\documentclass[12pt, A4]{article}
\usepackage[margin=1in]{geometry}
\usepackage[utf8]{inputenc}
\usepackage{comment}
\usepackage{ragged2e}
\usepackage{amsmath}
\usepackage{natbib}
\usepackage{array}
\usepackage{makecell} 
\usepackage{graphicx}
\usepackage[flushleft]{threeparttable}
\usepackage{float}
\usepackage{caption}
\usepackage{booktabs}
\usepackage{xcolor}
\usepackage{colortbl}
\usepackage{verbatim}
\usepackage{chngpage}
\usepackage{amsfonts}
\usepackage{enumitem}
\usepackage{setspace}
\allowdisplaybreaks[1]
\DeclareMathOperator{\vech}{vech}

\DeclareMathOperator{\Iden}{I}
\DeclareMathOperator{\Exp}{E}
\DeclareMathOperator{\trace}{trace}
\DeclareMathOperator{\var}{Var}
\DeclareMathOperator{\cov}{Cov}
\DeclareMathOperator{\diag}{diag}
\DeclareMathOperator{\st}{s.t.}
\DeclareMathAlphabet{\mathmybb}{U}{bbold}{m}{n}
\newcommand{\1}{\mathmybb{1}}
\newcommand\ph{$\phantom{1}$}
\font\myfont=cmr12 at 18pt
\begin{document}
\providecommand{\keywords}[1]
{{\textit{Keywords:}} #1}

\begin{singlespace}
\title{\myfont \textbf{Hierarchical DCC-HEAVY Model for High-Dimensional Covariance Matrices}\footnote{We are grateful to Fulvio Corsi (University of Pisa) for insightful comments and suggestions. We also thank Fulvio Corsi and Giuseppe Buccheri (University of Verona) for providing the HF dataset used in the paper, who acknowledge financial support from the PRIN Projects 20205J2WZ4, 2022MRSYB7 funded by the Italian MUR.}\vspace{0.8em}}
\author{Emilija Dzuverovic$^{a}$\footnote{Corresponding author. Department of Economics and Management, University of Pisa, Via Cosimo Ridolfi 10, 56124 Pisa, Italy. Email: emilija.dzuverovic@ec.unipi.it}, Matteo Barigozzi$^{b}$\vspace{0.5cm}\\
        \small $^{a}$ Department of Economics and Management, University of Pisa \\
        \small $^{b}$ Department of Economics, University of Bologna\\
\vspace{-0.2cm}}
\date{July 2024}
\maketitle
\end{singlespace}

\begin{abstract}
We introduce a HD DCC-HEAVY class of hierarchical-type factor models for high-dimensional covariance matrices, employing the realized measures built from higher-frequency data. The modelling approach features straightforward estimation and forecasting schemes, independent of the cross-sectional dimension of the assets under consideration, and accounts for sophisticated asymmetric dynamics in the covariances. Empirical analyses suggest that the HD DCC-HEAVY models have a better in-sample fit and deliver statistically and economically significant out-of-sample gains relative to the existing hierarchical factor model and standard benchmarks. The results are robust under different frequencies and market conditions.

\vspace{1.7em}
\hspace{-2cm}\textbf{Keywords}: {DCC-HEAVY; Asymmetric Volatility; Time-Varying Beta; High-Frequency Data.}

\hspace{-2cm}\textbf{JEL codes}:
\vspace{-5cm}
\thispagestyle{empty}
\graphicspath{{Graphs/}}
\end{abstract}
\clearpage
\pagenumbering{arabic} 
\maketitle

\newpage
\section{Introduction}
\justifying
Forecasting the covariance matrix of asset returns is paramount in asset pricing, portfolio selection, and risk management. In this regard, the majority of introduced models assume that variances and covariances are assessable conditional upon past daily information, i.e., multivariate GARCH (MGARCH) models (\cite{bauwens2006multivariate}), among which the DCC model by \cite{engle2002dynamic} is one of the most popular. In contrast, the increasing availability of intraday data has led to the development of so-called HEAVY models (e.g., \cite{shephard2010realising}, \cite{noureldin2012multivariate}, etc.), where the conditional covariances of daily returns are specified as a function of lagged realized covariances (RC). HEAVY models are based on more accurate measurements of covariances, thus improving the conditional forecasts of daily return co-movements (\cite{noureldin2012multivariate}).

In this paper, we introduce a class of High-Dimensional DCC-HEAVY (HD DCC-HEAVY) models for capturing the dynamics of multivariate financial time series, which allows for sophisticated asymmetric dynamics in the covariances based on the signs of underlying returns and, at the same time, keeps the estimation and forecasting straightforward and independent from the cross-sectional dimension of the assets under consideration.

Our methodology builds upon the Realized Beta GARCH model of \cite{hansen2014realized}, which introduces a hierarchical-type factor modelling framework to avoid the curse of dimensionality and estimation issues of widely applied covariance models. In particular, we first  build the marginal model for the set of selected observable factors via an extended DCC-HEAVY model (\cite{BauwensXu2022HEAVY}). Second, by assuming that the cross-correlation asset dynamics are largely driven by these factors, we specify the asymmetric HEAVY-type models for individual asset returns conditional on the factor estimates. As such, we explicitly model the dynamics of both conditional and RC based on the asymmetric DCC-HEAVY extension. In contrast, the approach of \cite{hansen2014realized} is built upon the Realized GARCH model (\cite{hansen2012realized}), thus taking the realized measures as direct inputs. Compared to the latter, multi-step ahead forecasts are available in closed form within our framework. Finally, we propose an additional third step to capture the idiosyncratic covariances via the nonlinear shrinkage of \cite{ledoit2017nonlinear}. 

The hierarchical structure of the model allows the adoption of a step-by-step estimation method in which first the parameters of the variance and covariance dynamics of the factors are separately estimated, second the parameters of the variance and covariance dynamics of each individual asset with factors are estimated and finally the idiosyncratic variance-covariance matrix is estimated on the model residuals. As such, the estimation method allows to apply the proposed model for an arbitrary large set of assets.

Given that no prior study investigates the forecasting ability of the hierarchical-type factor models, we assess the statistical and economic performance of the proposed model family, which incorporates distinct underlying factor sets with/out the focus on asymmetric effects, alongside the benchmark cDCC model and the Realized Beta GARCH model (\cite{hansen2014realized}). In addition to modelling the dynamics of daily returns and adopting intra-daily realized measures, we also use monthly returns and construct realized measures via daily data in order to test the models for much longer periods.

To perform empirical evaluations of the models at the daily vs. monthly frequency, we utilize the data from a Kenneth French library on the three Fama-French (FF) factors (\cite{fama1993common}), i.e., market risk, size, and value, together with the momentum factor (\cite{carhart1997persistence}), coupled with Yahoo Finance time series of the daily and monthly adjusted prices for a selected cross-section of individual assets, including all the stocks that belong to the S\&P500 Index during the entire sample period from January 1962 until January 2023, i.e., \(T = 732\). Concerning the intraday vs. daily analyses from January 2, 2014, to December 30, 2022, i.e., \(T=2266\), we adopt the intra-daily series on the same set of individual assets and build the high-frequency (HF) returns and corresponding realized measures for the analogous factor set based on the intraday exchange traded fund (ETF) data.

Both the in-sample and forecasting results imply that our HD DCC-HEAVY class of models significantly outperforms the existing hierarchical model of \cite{hansen2014realized} and the benchmark cDCC model. With regard to the latter, we confirm the benefits of employing the higher-frequency data to model conditional covariances of lower-frequency returns. Conversely, the importance of an employed DCC structure (see e.g., \cite{corsi2021dcc}) and/or the rich asymmetric dynamics could explain the inferior performance of the Realized Beta GARCH model (\cite{hansen2014realized}). The robustness of our findings is confirmed under each analyzed frequency and distinct market conditions.

The rest of the paper is organized as follows. Section 2 introduces the hierarchical HD DCC-HEAVY models. Sections 3 and 4 expound on the estimation and forecasting schemes. Section 5 describes the empirical methodology, details the data used in the paper, and presents the in- and out-of-sample results of the empirical exercises. Section 6 concludes. An additional summary of the data inputs and supplementary results of the empirical applications are presented in the appendices.

\section{Modelling Framework}
\justifying
Let us consider individual asset returns \(r_{t}^{i}\) (\(i=1,\dots,N\)) and a \(K \times 1\) vector \(r_{t}^{c}\) of close-to-close returns related to the set of \(K\) observable factors on day \(t\). In addition, \(RC_{t}^{c}\) and \(RC_{t}^{c,i}\) denote the realized covariance (RC) matrices of factors and an individual asset \(i\) vs. the factor set, respectively.

In this regard, we observe the two types of information sets, i.e., \({\cal F}_{t}^{c}\), composed of the variables related to the set of factors, and \({\cal F}_{t}^{c,i}\), which further incorporates the observable information on each individual asset, i.e., ${\cal F}_{t}^{c,i}={\cal F}_{t}^{c}\cup {\cal F}_{t}^{i}$.

We consider the following model for an individual asset return:
\begin{equation} \label{1}
r_{t}^{i} = \alpha_{t}^{i} + (\beta_{t}^{i})'r_{t}^{c} + \varepsilon_{t}^{i};
\end{equation}  
\begin{equation} 
\begin{split} \label{2}
\beta_{t}^{i} & = \var(r_{t}^c|{\cal F}_{t-1}^{c})^{-1}\cov(r_{t}^{i},r_{t}^{c}|{\cal F}_{t-1}^{c,i})\\ & =(\diag(H_{t}^{c})^{1/2}R_{t}^{c}\diag(H_{t}^{c})^{1/2})^{-1}\diag(H_{t}^{c})^{1/2}\rho_{t}^{i}(h_{t}^{i})^{1/2}\\&=(\diag(H_{t}^{c})^{1/2})^{-1}(R_{t}^{c})^{-1}\rho_{t}^{i}(h_{t}^{i})^{1/2};\\
\alpha_{t}^{i} &= \mu^{i} - (\beta_{t}^{i})'\mu^{c},
\end{split}
\end{equation}
where \(\alpha_{t}^{i}\) and \(\varepsilon_{t}^{i}\) are the intercept and idiosyncratic return component related to \(r_{t}^{i}\), respectively, $\mu^i$ is the constant trend parameter and $\mu^c$ is a constant vector, and \(\beta_{t}^{i}\) is a \(K \times 1\) vector of asset betas; \(\diag(H_{t}^{c})\) is a \(K \times K\) diagonal matrix composed of the conditional variances of factors on day \(t\), \(R_{t}^{c}\) is the corresponding \(K \times K\) conditional correlation matrix, while \(h_{t}^{i}\) denotes the conditional variance of an asset \(i\) and \(\rho_{t}^{i}\) a \(K \times 1\) vector of conditional correlations between an asset and the factors on day \(t\).

The model for \textit{N} individual asset returns follows readily by writing \eqref{1} in vector form:
\begin{equation} \label{3}
r_{t} = \alpha_{t} + B_{t}r_{t}^{c} + \varepsilon_{t},
\end{equation}
where \(r_{t}\) is a \(N \times 1\) vector of returns of individual assets on day \(t\), \(\alpha_{t}\) and \(\varepsilon_{t}\) ($\Exp(\varepsilon_{t})=0$) are the corresponding \(N \times 1\) vectors of intercepts and idiosyncratic return components, respectively, and \(B_{t}\) is a \(N \times K\) matrix of asset betas.

Our object of interest is:
\begin{equation} \label{4}
\var(r_{t}|{\cal F}_{t-1}^{c,i}) = B_{t}\var(r_{t}^c|{\cal F}_{t-1}^{c})(B_{t})' + \Sigma_{t},
\end{equation}
with \(\Sigma_{t} = \Exp(\varepsilon_{t}\varepsilon_{t}'|{\cal F}_{t-1}^{c,i})\).

To model (\ref{4}), we utilize the hierarchical factor approach of \cite{hansen2014realized}. In particular, \({\cal F}_{t}^{c}\) serves to specify the dynamics of the factor set. Subsequently, conditional on the former estimates, we set up the framework for the dynamics between each individual asset and the factors based on \({\cal F}_{t}^{c,i}\). Ultimately, the nonlinear shrinkage method (\cite{ledoit2017nonlinear}) is used to estimate $\Sigma_t$, containing the covariances between idiosyncratic return components of individual assets. 

Summing up, to estimate \eqref{4}, i.e., $\var(r_{t}^c|{\cal F}_{t-1}^{c})$, we need to estimate $H_t^c$ and $R_t^c$ (Section \ref{sec21}), $h_t^i$ and $\rho_t^i$ (Section \ref{sec22}), and last $\Sigma_t$ (Section \ref{sec23}).

\subsection{Model for Factors}\label{sec21}
\justifying
We specify the marginal model for a set of factors by extending the recently introduced DCC-HEAVY model of \cite{BauwensXu2022HEAVY} to allow for flexible asymmetric dynamics in the covariance matrices. 

In particular, each \(K \times K\) conditional covariance matrix of \(K\) factors, i.e., \(\var(r_{t}^{c}|{\cal F}_{t-1}^{c})=H_{t}^{c}\), 
is decomposed as (see also \eqref{2}):
\begin{equation} \label{5}
H_{t}^{c} = \diag(H_{t}^{c})^{1/2}R_{t}^{c}\diag(H_{t}^{c})^{1/2},
\end{equation}
where \(\diag(H_{t}^{c})\) is a diagonal matrix of conditional variances of factors on day \(t\) and \(R_{t}^{c}\) is the corresponding \(K \times K\) conditional correlation matrix.

Letting $h_t^c=\diag(H_t^c)v$, where $v$ is a $K$-dimensional vector of ones, the dynamics of the conditional variances $h_t^c$, allowing for asymmetric effects based on the signs of the underlying daily returns, and correlations are specified as follows:
\begin{equation} \label{6}
h_{t}^{c} = w_{h} + A_{h}^{+}v_{t-1}^{c} \odot \Iden_{t-1}^{+} + A_{h}^{-}v_{t-1}^{c} \odot \Iden_{t-1}^{-} + B_{h}h_{t-1}^{c},
\end{equation}
where \(v_{t}^{c}\) is the \(K \times 1\) vector of realized variances of factors on day \(t\),  \(w_{h}\) is a \(K \times 1\) positive vector, and \(A_{h}^{+}, A_{h}^{-}\), and \(B_{h}\) are  \(K \times K\) diagonal matrices of coefficients; \(\odot\) denotes the Hadamard (element-wise) product of matrices, and  \(\Iden_{t}^+ = [1_{\{r_{1,t}^{c} > 0\}}, \dots, 1_{\{r_{K,t}^{c} > 0\}}]'\) and \(\Iden_{t}^- = [1_{\{r_{1,t}^{c} \leq 0\}}, \dots, 1_{\{r_{K,t}^{c} \leq 0\}}]'\) the indicator vectors of the positive and negative daily returns, respectively.

Correspondingly, the conditional correlation matrix is given by:
\begin{equation} \label{7}
R_{t}^{c} = \tilde{R} + \alpha_{R}L_{t-1}^{c} + \beta_{R}R_{t-1}^{c},
\end{equation}
where \(L_{t}^{c}\) is a \(K \times K\) realized correlation matrix of the factors on day \(t\), \(\alpha_{R}\) and \(\beta_{R}\) are non-negative scalar parameter, and the targeting \(\tilde{R} =  (1 - \beta_{R})\overline{R} - \alpha_{R}\overline{P}\), with \(\overline{R}=\frac{1}{T}\sum_{t=1}^T R_t\) and \(\overline{P}=\frac{1}{T}\sum_{t=1}^TP_t\).

To model realized variances $v_t^c$ and correlations $L_t^c$ of factors on day \(t\), we decompose the \(K \times K\) conditional mean of the realized covariance (RC) matrix of the factors, i.e., \(\Exp(RC_{t}^{c}|{\cal F}_{t-1}^{c})=M_t^c\), as:
\begin{equation} \label{8}
M_t^c= \diag(M_{t}^{c})^{1/2}P_{t}^{c}\diag(M_{t}^{c})^{1/2},
\end{equation}
where \(\diag(M_{t}^{c})=\Exp(v_t^c| {\cal F}_{t-1}^c)\) is a \(K \times 1\) vector of the conditional means of realized variances of factors on day \(t\) and \(P_{t}^{c}=\Exp(L_{t}^{c}|{\cal F}_{t-1}^{c})\)
 is the corresponding \(K \times K\) conditional mean of realized correlations.

Letting $m_t^c=\diag(M_t^c)v$, the dynamics of the realized variances $m_t^c$ that accommodate the intra-daily asymmetric effects and correlations are specified as follows:
\begin{equation} \label{9}
m_{t}^{c} = w_{m} + A_{m}^{+}v_{t-1}^{c+} + A_{m}^{-}v_{t-1}^{c-} + B_{m}m_{t-1}^{c},
\end{equation}
where \(v_{t}^{c+}\) and \(v_{t}^{c-}\) are the \(K \times 1\) vectors of the positive and negative realized semi-variances (\cite{shephard2010realising}) of factors, respectively, \(w_{m}\) is a \(K \times 1\) positive vector, and \(A_{m}^{+}, A_{m}^{-}\), and \(B_{m}\) are  \(K \times K\) diagonal matrices of coefficients.
Specifically, for \(j=1, \dots, K\) and \(m=1, \dots, M\), \(v_{j,t}^{c+} =  \sum_{m=1}^{M}(r_{m,j,t}^{c+})^2\) and \(v_{j,t}^{c-} = \sum_{m=1}^{M}(r_{m,j,t}^{c-})^2\),
where \(r_{m,j,t}^{c+} = r_{m,j,t}^{c} \times 1_{\{r_{m,j,t}^{c} > 0\}}\) and \(r_{m,j,t}^{c-} = r_{m,j,t}^{c} \times 1_{\{r_{m,j,t}^{c} \leq 0\}}\) denote the positive and negative intraday returns, respectively.

Finally,
\begin{equation} \label{10}
P_{t}^{c} = (1 - \alpha_{P} - \beta_{P})\overline{P} + \alpha_{P}L_{t-1}^{c} + \beta_{R}P_{t-1}^{c},
\end{equation}
where \(\alpha_{P}\) and \(\beta_{P}\) are non-negative scalar parameters, and \(\overline{P}=\frac{1}{T}\sum_{t=1}^TP_t\).

\subsection{Model for Individual Asset Returns}\label{sec22}
\justifying
Given the assumption of the conditional dependence of the distribution of individual asset returns on the factors but not vice versa (\cite{hansen2014realized}), operating through the conditional correlations, the \((K + 1) \times (K + 1)\) joint conditional correlation matrix \(R_{t}^{c,i}=\text{Corr}([r_t^c\ r_t^i])\) is given by:
\begin{equation} \label{11}
R_{t}^{c,i}=
\begin{pmatrix}
R_{t}^{c} & \rho_{t}^{i}\\
(\rho_{t}^{i})' & 1
\end{pmatrix},
\end{equation}
where \(R_{t}^{c}\) and \(\rho_{t}^{i}\) denote the \(K \times K\) conditional correlation matrix of factors filtered from a marginal model and \(K \times 1\)  vector of correlations between an individual asset and the factors on day \(t\), respectively. The law of motion of $R_t^c$ is defined in \eqref{7} while the one for $\rho_{t}^{i}$ is defined through \eqref{14} below. 

In accordance with the framework for a set of factors, the dynamics of the conditional and realized variance of each individual asset, allowing for asymmetric effects based on corresponding signed returns, are specified as follows:
\begin{equation} \label{12}
h_{t}^{i} = c_{h}^{i} + a_{h}^{i+}v_{t-1}^{i} 1_{[r_{t-1}^{i}>0]} + a_{h}^{i-}v_{t-1}^{i} 1_{[r_{t-1}^{i}\leq 0]} + b_{h}^{i}h_{t-1}^{i},
\end{equation}
where \(h_{t}^{i}\)\ and \(v_{t}^{i}\) denote the conditional and realized variance of an asset \(i\) on day \(t\), respectively, and \(c_{h}^{i}, a_{h}^{i+}, a_{h}^{i-}\), and \(b_{h}^{i}\) are non-negative scalar coefficients;
\begin{equation} \label{13}
m_{t}^{i} = c_{m}^{i} + a_{m}^{i+}v_{t-1}^{i+} + a_{m}^{i-}v_{t-1}^{i-} + b_{m}^{i}m_{t-1}^{i},
\end{equation}
where \(m_{t}^{i}\), \(v_{t}^{i+}\), and \(v_{t}^{i-}\) denote the conditional mean of the realized variance, positive and negative semi-variance of an asset \(i\) on day \(t\), respectively, and \(c_{m}^{i}, a_{m}^{i+}, a_{m}^{i-}\), and \(b_{m}^{i}\) are non-negative scalar coefficients.

Finally, to model the vectors of correlations between the returns of an individual asset and the set of factors, we utilize the Fisher transformation, i.e., \(\mathbb{F}(\cdot)\), mapping each element from a closed interval \((-1,1)\) into \(\mathbb{R}\), within the typical HEAVY-type recursions (\cite{noureldin2012multivariate}, \cite{BauwensXu2022HEAVY}): 
\begin{equation} \label{14}
\mathbb{F}(\rho_{t}^{i}) = \phi_{R}^{i} + \alpha_{R}^{i}\mathbb{F}(l_{t-1}^{i}) + \beta_{R}^{i}\mathbb{F}(\rho_{t-1}^{i}),
\end{equation}
where \(\rho_{t}^{i}\) and \(l_{t}^{i}\) denote the \(K \times 1\) vectors of conditional and realized correlations of an asset \(i\) with factors on day \(t\), respectively, and \(\phi_{R}^{i}\), \(\alpha_{R}^{i}\), and \(\beta_{R}^{i}\) are non-negative scalar parameters;
\begin{equation} \label{15}
\mathbb{F}(p_{t}^{i}) = \phi_{P}^{i} + \alpha_{P}^{i}\mathbb{F}(l_{t-1}^{i}) + \beta_{P}^{i}\mathbb{F}(p_{t-1}^{i}),
\end{equation}
where \(p_{t}^{i}\) denotes a \(K \times 1\) vector of the conditional means of realized correlations of an asset \(i\) with factors on day \(t\), and \(\phi_{P}^{i}\), \(\alpha_{P}^{i}\), and \(\beta_{P}^{i}\) are non-negative scalar parameters.

\subsection{Idiosyncratic Dynamics}\label{sec23}
Based on formulas (\ref{1})–(\ref{4}), to fully specify the conditional covariance matrices of individual assets, we should define the dynamics of the conditional covariance of the residuals, i.e., \(\Sigma_t=\Exp(\varepsilon_{t}\varepsilon_{t}'|{\cal F}_{t-1}^{c,i})\). 

In principle, the dynamic of \(\Sigma_t\) could be determined, e.g., via the benchmark dynamic conditional correlation (DCC) model (\cite{engle2002dynamic}) for the cross-section of not to many assets, e.g., \(N \leq 100\). Otherwise, when the number of individual assets is large, the DCC-NL model introduced by \cite{engle2019large}, i.e., DCC with correlation targeting estimated by the nonlinear shrinkage estimator, might be adopted. In each case, the estimation of at least additional \(3N + 2\) parameters is required. Thus, to maintain the model parsimony, we assume the conditional covariances of the residuals to be constant in time, i.e., $\Sigma_t=\Sigma_\epsilon$.

In line with most of the literature, we consider the assumption of an exact factor model as too strict, hence we do not assume $\Sigma_\epsilon$ to be diagonal. Therefore, given an underlying approximate factor model, we propose applying the nonlinear shrinkage method of \cite{ledoit2017nonlinear} to the sample covariance matrix of the residuals, which has been proven preferable with respect to both the linear shrinkage of \cite{ledoit2004well} (\cite{ledoit2017nonlinear}) and thresholding schemes (\cite{de2021factor}). The selected methodology implies shifting the eigenvalues of the empirical covariance matrix via the out-of-sample optimization of the minimum variance loss function subject to a required return constraint (\cite{engle2006testing}).


\section{Estimation}
\justifying
The hierarchical structure of the introduced model suggests a convenient step-by-step quasi-maximum likelihood (QML) estimation scheme that can be applied to any cross-sectional dimension of the assets under consideration.

Initially, to estimate the core model for a set of factors, we follow the approach of \cite{BauwensXu2022HEAVY} and partition the vectors of parameters \(\theta_H^c\) and \(\theta_M^c\) for the conditional and realized covariances of factors, respectively, in two sets, containing the coefficients of the corresponding variance and correlation equations.\footnote{The parameters can be alternatively estimated jointly with respect to the corresponding LLFs (see \cite{BauwensXu2022HEAVY}).}

The first step consists of estimating the parameters of the conditional variances (\ref{6}), i.e., \(\theta_{H_1}^c\), and correlations (\ref{7}), i.e., \(\theta_{H_2}^c\), for the set of factors by maximizing
the following QML functions:
\setcounter{equation}{17}
\begin{equation}  \label{18}
\begin{split}
LLF_{H_1}^c(\theta_{H_1}^c|{\cal F}_{t-1}^{c})& =-\frac{1}{2}\sum_{t=1}^T \left\{2\log \left|\diag(H_{t}^{c})^{1/2}\right|+ u_{t}^{c'}u_{t}^{c}\right\};\\
LLF_{H_2}^c(\theta_{H_2}^c|\hat{\theta}_{H_1}^c;{\cal F}_{t-1}^{c})& =-\frac{1}{2}\sum_{t=1}^T \left\{\log \left|R_{t}^{c}\right|+ \hat{u}_{t}^{c'}(R_{t}^{c})^{-1}\hat{u}_{t}^{c}\right\},
\end{split}
\end{equation}
where \(\hat{u}_{t}^{c} = r_{t}^{c} \odot (\hat{h}_{t}^{c})^{-1/2}\), with the vector of conditional variances \(\hat{h}_{t}^{c}\) defined via \(\hat{\theta}_{H_1}^c\).\footnote{\cite{BauwensXu2022HEAVY} show that the estimated parameters for conditional correlations (\ref{7}), i.e., \((\hat{\alpha}_{R}, \hat{\beta}_{R})\), do not automatically guarantee the PD-ness of \(\hat{R}_{t}^{c}\). As such, we proceed by checking the condition during the numerical maximization of \(LLF_{H_2}^c\).}

To specify the dynamics of realized measures, we assume that their conditional probability density function is Wishart, i.e.,
\begin{equation} \label{19}
RC_t^c|{\cal F}_{t-1}^c \sim W_K(\nu, M_t^c(\theta_M^c)/\nu), 
\end{equation}
where \(W_K(\nu, M_t^c(\theta_M^c)/\nu)\) denotes the \(K\)-dimensional central Wishart distribution with $\nu \geq K$ degrees of freedom and positive-definite (PD) \(K \times K\) scale matrix \(M_t^c(\theta_M^c)/\nu\), implying \(\Exp(RC_t^c|{\cal F}_{t-1}^c) = M_t^c(\theta_M^c)\).

Correspondingly, we split \(\theta_M^c\) into the parameters for realized variances (\ref{9}), i.e., \(\theta_{M_1}^c\), and realized correlations (\ref{10}), i.e., \(\theta_{M_2}^c\). The second-step objective functions for \(T\) observations are given by:\\
\begin{equation}  \label{20}
\begin{split}
LLF_{M_1}^c(\theta_{M_1}^c|{\cal F}_{t-1}^{c}) =& -\frac{\nu}{2}\sum_{t=1}^T \left\{2\log \left|S_{m,t}^c\right| + \trace\left[(S_{m,t}^c)^{-1}RC_t^c(S_{m,t}^c)^{-1}\right]\right\};\\
LLF_{M_2}^c(\theta_{M_2}^c|\hat{\theta}_{M_1}^c;{\cal F}_{t-1}^{c}) =& -\frac{\nu}{2}\sum_{t=1}^T \left\{\log \left|P_{t}^{c}\right|+ \trace\left[((P_t^{c})^{-1} - \Iden_K)(\hat{S}_{m,t}^c)^{-1}RC_t^c(\hat{S}_{m,t}^c)^{-1}\right]\right\},
\end{split}
\end{equation}
where \(\Iden_K\) denotes the identity matrix of order \(K\), \(S_{m,t}^c = \diag(M_{t}^{c})^{1/2}\), with \(\hat{S}_{m,t}^{c}\) defined via \(\hat{\theta}_{M_1}^c\), and the parameter \(\nu\) set equal to 1.\footnote{The score for \(\theta_M^c\) is proportional to \(\nu\).}

Next, we consider the likelihood contributions for the model for individual assets. Given the assumptions of the conditional dependence on the factors, the standardized daily asset return satisfies:
\begin{equation}  \label{21}
u_{t}^{i}|u_{t}^{c}  \sim N \left( (\rho_{t}^{i})'(R_{t}^{c})^{-1}u_{t}^{c}, 1 - (\rho_{t}^{i})'(R_{t}^{c})^{-1}\rho_{t}^{i}\right).
\end{equation}

As such, the underlying LLF with regard to the conditional covariances of an asset \(i\)
\begin{equation}  \label{22}
\begin{split}
LLF_{H_i}^{c,i}(\theta_{H_i}|{\cal F}_{t-1}^{c,i}) =-\frac{1}{2}\sum_{t=1}^T \left\{\log \left(h_{t}^{i}\left(1 - (\rho_{t}^{i})'(R_{t}^{c})^{-1}\rho_{t}^{i}\right)\right)+ \frac{(u_{t}^{i}-(\rho_{t}^{i})'(R_{t}^{c})^{-1}u_{t}^{c})^2}{\left(1 - (\rho_{t}^{i})'(R_{t}^{c})^{-1}\rho_{t}^{i}\right)}\right\},
\vspace{-1cm}
\end{split}
\end{equation}

directly follows from:
\begin{equation}  \label{23}
\begin{split}
\cov(r_{t}^{i},r_{t}^{c}|{\cal F}_{t-1}^{c,i})&=\diag(h_{t}^{c})^{1/2}\rho_{t}^{i}(h_{t}^{i})^{1/2};\\
\var(r_{t}^{i}|r_{t}^{c},{\cal F}_{t-1}^{c,i})&=h_{t}^{i} - \frac{(\diag(h_{t}^{c})^{1/2}\rho_{t}^{i}(h_{t}^{i})^{1/2})'(\diag(h_{t}^{c})^{1/2}\rho_{t}^{i}(h_{t}^{i})^{1/2})}{\diag(h_{t}^{c})^{1/2}R_{t}^{c}\diag(h_{t}^{c})^{1/2}}\\& = h_{t}^{i}\left(1 - (\rho_{t}^{i})'(R_{t}^{c})^{-1}\rho_{t}^{i}\right);\\
\Exp(r_{t}^{i}|r_{t}^{c},{\cal F}_{t-1}^{c,i})&=\mu^{i} + \frac{(\diag(h_{t}^{c})^{1/2}\rho_{t}^{i}(h_{t}^{i})^{1/2})'}{\diag(h_{t}^{c})^{1/2}R_{t}^{c}\diag(h_{t}^{c})^{1/2}}(r_{t}^{c} - \mu^{c}) = \mu^{i} + (h_{t}^{i})^{1/2}(\rho_{t}^{i})'(R_{t}^{c})^{-1}u_{t}^{c}.
\end{split}
\end{equation}

In analogous fashion as for the conditional correlations (\ref{11}), we use a partitioning of the realized measures so that, e.g., the \((K + 1) \times (K + 1)\) joint conditional mean of the realized correlation matrix \(P_{t}^{c,i}\) is given by:
\begin{equation} \label{24}
P_{t}^{c,i}=
\begin{pmatrix}
P_{t}^{c} & p_{t}^{i}\\
(p_{t}^{i})' & 1
\end{pmatrix},
\end{equation}
where \(P_{t}^{c}\) and \(p_{t}^{i}\) denote the \(K \times K\) conditional mean of the realized correlation matrix of factors filtered from a marginal model and \(K \times 1\)  vector of the conditional expectations of correlations between an individual asset and the factors on day \(t\), respectively.

In this regard, the QML function reads as:\\
\begin{equation}  \label{25}
\begin{split}
LLF_{M_i}^{c,i}(\theta_{M_{i}}|{\cal F}_{t-1}^{c,i}) = -\frac{\nu}{2}\sum_{t=1}^T \left\{\log\left(m_{t}^{i}(1-p_{t}^{i|c})\right) + \frac{v_{t}^{i}-(rc_{t}^{i})'(RC_{t}^{c})^{-1}rc_{t}^{i}}{m_{t}^{i}(1-p_{t}^{i|c})}\right\},\\
\end{split}
\end{equation}
where \(m_{t}^{i}\) denotes the conditional mean of the realized variance of an asset \(i\), i.e., \(v_{t}^{i}\), \(rc_{t}^{i}\) is a \(K \times 1\) vector of the realized covariances between an asset \(i\) and \(K\) factors, and \(p_{t}^{i|c} = (p_{t}^{i})'(P_{t}^{c})^{-1}p_{t}^{i}\). Analogously, we set \(\nu\) equal to 1.
\newpage
\subsection{Summary of Estimation}
Summarizing, the proposed step-by-step estimation method applied to an arbitrary large cross-section of $N$ assets is computed by performing the following steps: 
\vspace{-0.1cm}
\begin{enumerate}[label=\textbf{\theenumi}]
\item marginal model for a set of factors via \eqref{18} and \eqref{20}, i.e.,
\begin{enumerate}[label=\textbf{\theenumi\alph*}]
\item estimate the parameters in the dynamics of the factor conditional variances and correlations via (\ref{6}) and (\ref{7}), 
\item estimate the parameters in the dynamics of the factor RV and realized correlations via (\ref{9}) and (\ref{10});
\end{enumerate}
\item individual models for \(i = 1, ...,N\) assets via \eqref{22} and \eqref{25}, i.e.,
\begin{enumerate}[label=\textbf{\theenumi\alph*}]
\item estimate the parameters in the dynamics of the asset conditional variance and correlations with factors via (\ref{12}) and (\ref{14}), 
\item estimate the parameters in the dynamics of the asset RV and realized correlations with factors via (\ref{13}) and (\ref{15});
\end{enumerate}
\item estimate covariances of residuals with nonlinear shrinkage (\cite{ledoit2017nonlinear}).
\end{enumerate}

Considering the estimation of the core model, the total number of parameters with respect to \(K\) factors is \(8K+4\). Given the assumption of diagonal matrices of coefficients for the variance equations, we split the estimation of \(8K\) parameters for the variances into \(K\) univariate HEAVY models (\cite{shephard2010realising}). Correspondingly, the model for each individual asset requires the specification of 14 additional parameters. As follows, a total of \(8K + 4 + 14N\) coefficients is estimated via the step-by-step QML, independent of both \(K\) and \(N\).

\section{Forecasting}
\justifying
Within the proposed framework, we focus on the 1-step-ahead predictions of the daily return covariances for the selected cross-section of \(N\) individual assets, i.e., \(\var(r_{t+1}|{\cal F}_{t}^{c,i})\), directly computable via:
\begin{equation} \label{26}
\widehat{\var}(r_{t+1}|{\cal F}_{t}^{c,i}) = \hat{B}_{t+1}\widehat{\var}(r_{t+1}^{c}|{\cal F}_{t}^{c})(\hat{B}_{t+1})'+ \hat{\Sigma}_{\hat{\varepsilon}} = \hat{B}_{t+1}\hat{H}_{t+1}^{c}(\hat{B}_{t+1})' + \hat{\Sigma}_{\hat{\varepsilon}},
\end{equation}
where \(\hat{H}_{t+1}^{c}\) is a \(K \times K\) predicted conditional covariance matrix of factors for the next day, \(\hat{B}_{t+1}\) is a \(N \times K\) matrix of predicted asset betas, and \(\hat{\Sigma}_{\hat{\varepsilon}}\) is a \(N \times N\) covariance matrix of the forecasted residuals.

In particular, for each asset \(i\) and time \(t+1\): 
\begin{equation} 
\begin{split} \label{27}
\hat{\beta}_{t+1}^{i} &=(\diag(\hat{H}_{t+1}^{c})^{1/2})^{-1}(\hat{R}_{t+1}^{c})^{-1}\hat{\rho}_{t+1}^{i}(\hat{h}_{t+1}^{i})^{1/2},
\end{split}
\end{equation}
where \(\diag(\hat{H}_{t+1}^{c})\) is a \(K \times K\) diagonal matrix composed of the conditional variances of factors for the day \(t+1\), \(\hat{R}_{t+1}^{c}\)
is the corresponding \(K \times K\) conditional correlation matrix. \(\hat{h}_{t+1}^{i}\) denotes the predicted conditional variance of an asset \(i\), and \(\hat{\rho}_{t+1}^{i}\) a \(K \times 1\) vector of the forecasted conditional correlations between an asset \(i\) and the factors.

\section{Empirical Application}
\subsection{Data Construction and Description}
\justifying
For the subsequent empirical analyses, we compute daily covariance matrices with the corresponding realized analogues with respect to the three Fama-French (FF) factors (\cite{fama1993common}), i.e., market risk, size, and value, together with the momentum factor (\cite{carhart1997persistence}). To accurately replicate the selected factor set intra-daily, we use an approach based on HF ETF data obtained from Kibot (see, e.g., \cite{bannouh2012realized}).\footnote{Alternatevily, the intra-daily factor series can be generated using the HF data with respect to the entire universe of stocks listed on NYSE, NASDAQ, and AMEX (see, e.g., \cite{ait2020high}).}

We consider the equity ETFs from the major U.S. investment management companies, i.e., BlackRock and Vanguard Group. In particular, we use the Vanguard Total Stock Market ETF (VTI) to replicate the market portfolio, which has the superior trading intensity compared to BlackRock's IWV. On the other hand, the iShares Russell funds that track the performance of small and large-cap firms, coupled with low, medium, and high book-to-market ratio ones, are selected to replicate the size and value factors, respectively.\footnote{Specifically, the data on IWB, IWD, IWF, IWM, IWN, and IWO.} These funds have both a sufficiently long and liquid trading history. Ultimately, for the momentum factor, we use the BlackRock MTUM, considering the trading volume during the selected sample period from January 2, 2014, to December 30, 2022, i.e., \(T=2266\), in order to account for low and high volatility cycles.

Initially, we determine the intraday log prices of the factors as linear combinations of selected fund prices, following the FF definitions (\cite{fama1993common, fama2015five}), via the refresh time sampling applied to the asynchronous series of all funds. Then, upon obtaining the intra-daily return vectors, we compute the RCs and their semi-decomposition (\cite{bollerslev2020realized}) using the simple benchmark estimator at the 5-min frequency. Finally, we calculate the RC matrices between an individual asset and the factor set using the same methodology with regard to all the stocks that belong to the S\&P500 Index during the period from January 1962 until January 2023, i.e., \(N = 19\).\footnote{All the empirical analyses at the daily vs. monthly frequency have been replicated within the above sample period, i.e., based on monthly covariance matrices and realized measures built upon daily data, with the summary results presented in Appendix C.} The stock names and tickers can be found in Appendix A. 

Table \ref{table:1} reports, for each factor, the time series mean and standard deviation of the realized variance (annualized in percentage, i.e., multiplied by 25200) and of the ‘positive’ and ‘negative’ components that are used to specify the asymmetric dynamics.\footnote{To avoid the information losses of the refresh time sampling applied to the asynchronous intraday data, we model each individual realized variance based on the corresponding (constructed) univariate intra-daily series.}. The last row indicates the average of the time series means and standard deviations of realized correlations between the factors. The same statistics for the individual assets are shown in Appendix \ref{AppA}, i.e., Table \ref{table:A1}.

\begin{table} [H] 
\begin{adjustwidth}{-.6in}{-.6in} \begin{center}
\scalebox{0.9}{
\begin{threeparttable}
\caption{Summary statistics of daily measures for the 3 FF and MOM factors} \label{table:1}
\begin{tabular}{l c c c c}
 \hline
Factor & MKT & SMB & HML & MOM \\
 \hline
\textit{r}$_{cc}^2$ & 3.38 (13.7) & 1.04 (2.31) & 1.04 (3.00) & 4.11 (14.9) \\
$RV$ & 1.89 (4.69) & 1.08 (2.49) & 0.82 (1.75) & 3.07 (28.0) \\
\hline
$P$ & 0.94 (2.58) & 0.54 (1.42) & 0.41 (0.83) & 1.49 (12.6) \\
$N$ & 0.95 (2.20) & 0.54 (1.20) & 0.41 (1.01) & 1.58 (15.4) \\
\hline
$GJR_{P}$ & 0.81 (2.85) & 0.49 (1.73) & 0.39 (0.93) & 1.08 (3.09) \\
$GJR_{N}$ & 1.08 (3.95) & 0.59 (1.95) & 0.43 (1.59) & 1.99 (27.9) \\
\hline
$RL$ & 0.41 (0.21) & 0.43 (0.22) & 0.15 (0.25) & 0.43 (0.20) \\
\hline
\end{tabular}
\begin{tablenotes}[flushleft]
\footnotesize 
\item 
$r_{cc}^2$: squared close-to-close return;
$RV$: realized variance; $P$: positive semi-variance; $N$: negative semi-variance;
$GJR_P$: $RV$ if daily return is positive, $0$ if negative; 
$GJR_N$:  $RV$ if daily return is negative, $0$ if positive;
$RL$:  realized correlation, the average of the 3 time series means and sd-s of realized correlations with the other 3 factors.
\end{tablenotes}
\end{threeparttable}}
\end{center}
\end{adjustwidth}
\end{table}

Considering the statistics reported in Table \ref{table:1}, the market and momentum factors appear more volatile compared to the size and value factors. For the former, the average negative semi-variance (\(N\)) is slightly larger than the average positive component (\(P\)). Conversely, the average portion of the variance with respect to the negative daily returns \(GJR_N\) exceeds the corresponding positive \(GJR_P\) for each factor. Besides HML, each factor is significantly correlated with the others.

The analogous summary measures for the individual assets, i.e., Table \ref{table:A1}, confirm that each average realized variance is only a fraction of the corresponding squared close-to-close return. It might not be surprising, given the daily return accounts for the overnight information. In general, the average positive semi-variance (\(P\)) is larger than the negative component (\(N\)). Conversely, as for the factor set, the portions of the variances with respect to the negative daily returns, i.e., \(GJR_N\), exceed \(GJR_P\). Ultimately, the average realized correlations of all the assets with the market factor are diverse, ranging from 0.17 to 0.57, with rather similar standard deviations.

\begin{figure}[H]
\centering
\includegraphics[height=10cm,width=\textwidth]{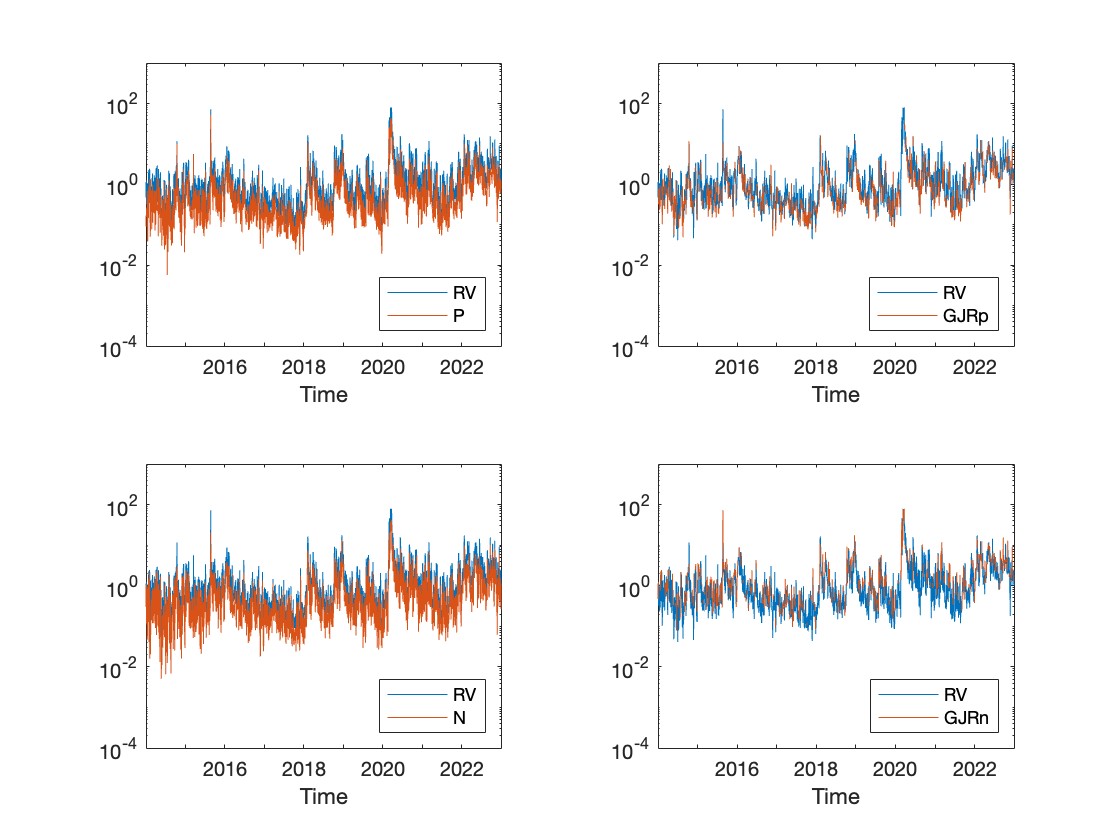}
\caption{Annualized log-transformed realized variance of the market factor and the terms of its decomposition using the signed intraday (left panel) and daily close-to-close returns (right panel)}\label{fig:1}
\end{figure}

\vspace{-1cm}
\begin{figure}[H]
\centering
\includegraphics[height=10cm,width=\textwidth]{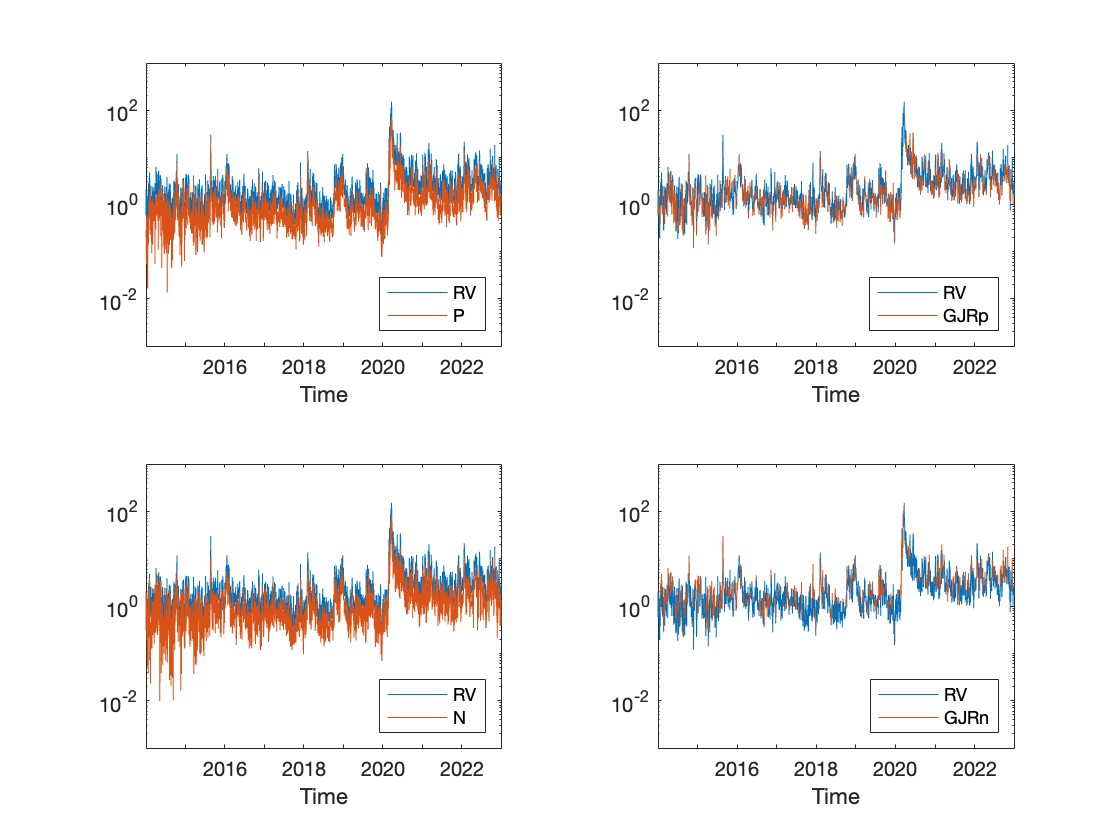}
\caption{Annualized log-transformed realized variance of the momentum factor and the terms of its decomposition using the signed intraday (left panel) and daily close-to-close returns (right panel)}\label{fig:2}
\end{figure}
\vspace{-0.5cm}
\begin{figure}[H]
\centering
\includegraphics[height=10cm,width=\textwidth]{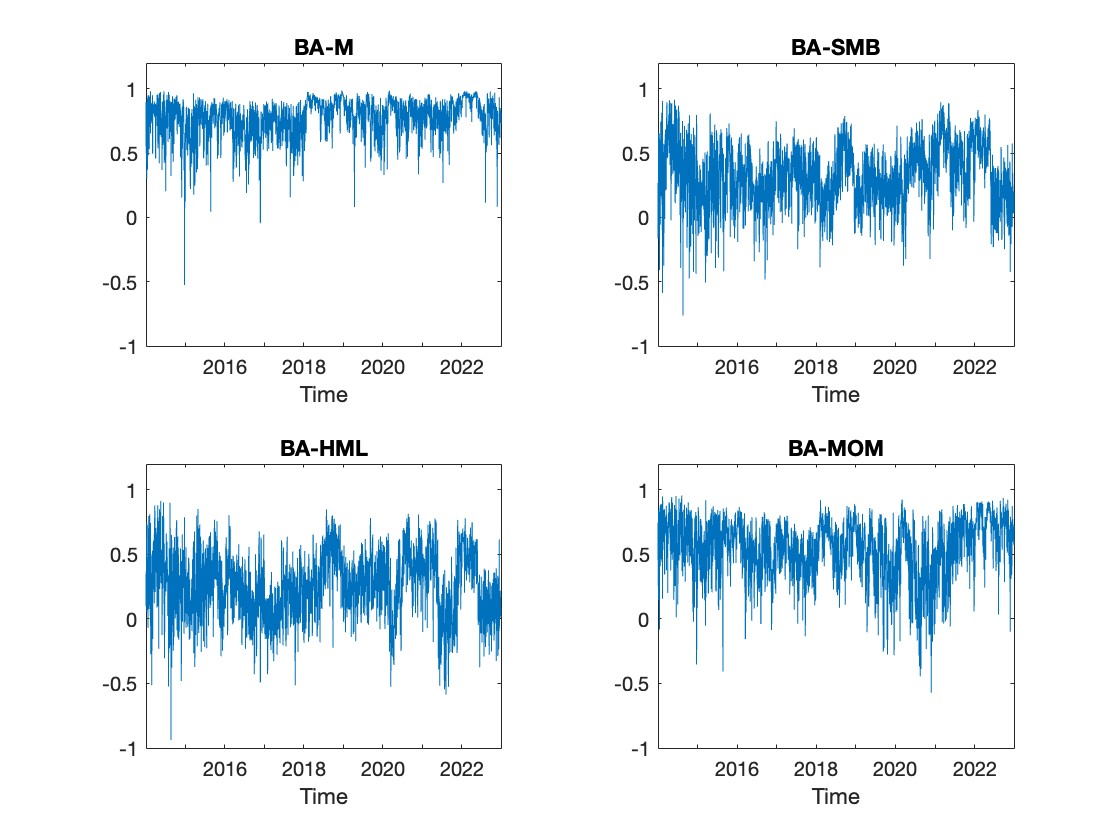}
\caption{Realized correlations of BA with the three FF factors and MOM}\label{fig:3}
\end{figure}

Figures \ref{fig:1} and \ref{fig:2} show the time series of the realized variance of the market and momentum factor, respectively, and the corresponding decomposition based on signed intraday and daily returns. They illustrate the occurrence of clustered extreme values consistent with periods of financial and economic turbulence. Ultimately, Figure \ref{fig:3} illustrates the time series of the realized correlations between BA and each factor. Unsurprisingly, BA is mostly correlated with the market factor, while the correlations with the other factors are more dispersed and volatile.

\subsection{In-Sample Fit}
\justifying
To evaluate the in-sample fit of our benchmark model, entitled the 4 Factor High-Dimensional DCC-HEAVY (“4F-HD DCC-HEAVY'') with \(K=4\), i.e., 3 FF factors and MOM, we additionally consider the restricted versions with respect to the set of factors and asymmetric effects.

In particular, we estimate the variants by assuming that equity returns are either explained via the 3 FF factors (“FF-HD DCC-HEAVY'') or the market factor only (“M-HD DCC-HEAVY''). In addition, we examine whether allowing for asymmetric effects in the covariance dynamics improves the fit by specifying the modelling equations of the benchmark model without accounting for the signs of underlying returns (“sym-HD DCC-HEAVY''). In this regard, we impose $A^+ = A^- = A$ on the variance equations for “sym-HD DCC-HEAVY''.

Correspondingly, to test the potential additional information provided by the HF data, we compare our model with the standard cDCC model built exclusively upon daily data, which has been widely applied to capture the dynamics of time-varying betas (e.g., \cite{bali2017dynamic}, \cite{engle2012dynamic}, etc.). Ultimately, we consider the existing hierarchical factor model, i.e., the Realized Beta GARCH model of  \cite{hansen2014realized}.

We estimate each model for a selected cross-section of indiidual assets. The in-sample fit of the six models has been assessed using the three criteria, i.e., the value of the maximized LLF, the Akaike information criterion (AIC), and the Bayesian information criterion (BIC). For all the models, we present the average value of each criterion with respect to \(N=19\) estimation results.\footnote{The full set of results is available upon request.} 

In view of the results presented in Table \ref{Table-LLF-AIC-BIC}, several conclusions can be drawn:

\begin{enumerate}
\item The “\textbf{4F-HD DCC-HEAVY}'' model has a larger LLF value and smaller AIC and BIC than the symmetric version “\textbf{sym-HD DCC-HEAVY}'' with respect to both core and conditional models for individual assets. As follows, allowing for the asymmetric dynamics in the covariances of factors, as well as an individual asset vs. the set of factors based on the signs of underlying (intra-)daily returns, improves the in-sample fit of the model.

\item Considering the fit criteria evaluated at conditional models for individual assets, the “\textbf{4F-HD DCC-HEAVY}'' model outperforms the FF and M versions. So, the additional factors increase the explanatory power of the model.

\item Among the market factor-based models, considering the total LLF values and both information criteria evaluated at the daily data, the best fitting model is “\textbf{M-HD DCC-HEAVY}''. The relative superiority of our model suggests the benefits of adopting higher-frequency data to model conditional covariances of lower-frequency returns as opposed to the \textbf{cDCC} model. Furthermore, the DCC-type structure and/or sophisticated asymmetric dynamics are important, as “\textbf{M-HD DCC-HEAVY}'' outperforms the \textbf{Realized Beta GARCH} model of \cite{hansen2014realized}. The latter provides for a better fit with respect to each criterion compared to the low-frequency data-based \textbf{cDCC} model.
\end{enumerate}

\begin{table}[H]
\begin{center}
\scalebox{0.7}{
\begin{threeparttable}
\caption{Maximum log-likelihood function (LLF), AIC, and BIC values of estimated models} \label{Table-LLF-AIC-BIC}
\begin{tabular}{l c c c c}
 \hline
& \textbf{4F-HD DCC-HEAVY} & \textbf{sym-HD DCC-HEAVY} & \textbf{FF-HD DCC-HEAVY} & \textbf{M-HD DCC-HEAVY}\\
\hline
LLF$^c$ & \textbf{-88894.72} & -89075.66 & / & /\\
AIC & \textbf{78.491} & 78.647 & / & /\\
BIC & \textbf{78.582} & 78.728 & / & /\\
& & & \\
LLF$^{c,i}$ & \textbf{-14258.36} & -15804.11 & -15772.94 & --18165.70\\
AIC & \textbf{12.597} & 13.959 & 13.934 & 16.046\\
BIC & \textbf{12.632} & 13.990 & 13.969 & 16.081\\
& & & \\
LLF$^c$ $+$ LLF$^{c,i}$ & \textbf{-103153.08} & -104879.77 & / & /\\
AIC & \textbf{91.088} & 92.606 & / & /\\
BIC & \textbf{91.215} & 92.718 & / & /\\
\hline
& \textbf{M-HD DCC-HEAVY} & \textbf{Real. Beta GARCH} & \textbf{cDCC} & \\
\hline
LLF$_{H}^{c}$ $+$ LLF$_{H_i}^{c,i}$ & \textbf{-18668.84} & -18679.95 & -20370.54 & \\
AIC & \textbf{16.487} & 16.500 & 17.986 & \\
BIC & \textbf{16.515} & 16.535 & 18.007 & \\
 \hline
\end{tabular}
\begin{tablenotes}[flushleft]
\footnotesize 
\item 
LLF\(^{c}\): total LLF for the core model;\\ LLF\(^{c,i}\): average (across \(N\) assets) total LLF for the conditional model for individual assets;\\ LLF$_{H}^{c}$ \(+\) LLF\(_{H_i}^{c,i}\): average (across \(N\) assets) total LLF evaluated at the daily data;\\
For each maximum value of the log-likelihood function (LLF), we report the corresponding Akaike (AIC) and Bayesian information criteria (BIC). The values in bold correspond to the best model of each row. The models are estimated using the dataset of 2266 observations described in Section 5.1.
\end{tablenotes}
\end{threeparttable}}
\end{center}
\end{table}

\vspace{-0.5cm}The estimates of the parameters of the core model for each HD DCC-HEAVY version are reported in Table \ref{Table:core model}. The results demonstrate that the coefficients in columns III-V notably differ for the “\textbf{M-HD DCC-HEAVY}'' model, implying distinct dynamics of the variance of the market factor compared to the others.

In each case, the average estimate of the \(b_h\) parameter is much smaller compared to standard GARCH models, while the average estimates of the \(a_h^+\) and \(a_h^-\) parameters are much larger compared to conventional ARCH terms. In line with the findings
of, e.g., \cite{BauwensXu2022HEAVY}, \cite{noureldin2012multivariate}, \cite{shephard2010realising},  etc., these results suggest that the dynamics of conditional variances are better captured by realized variances than by squared returns. Columns VI-VII present the parameter estimates of the correlations, implying persistent series typically found in the literature. 

In general, all the factors exhibit a significant leverage effect with respect to underlying (intra-)daily returns, i.e., the greater \(a_h^-\) and \(a_m^-\) parameters compared to \(a_h^+\) and \(a_m^+\), respectively. Thus, one of the main stylized facts of the financial return series, i.e., the stronger impact of negative returns on volatility, seems incorporated in the dynamics of the ETF-based portfolio returns.

\begin{figure}[H]
\centering
\includegraphics[height=9cm,width=\textwidth]{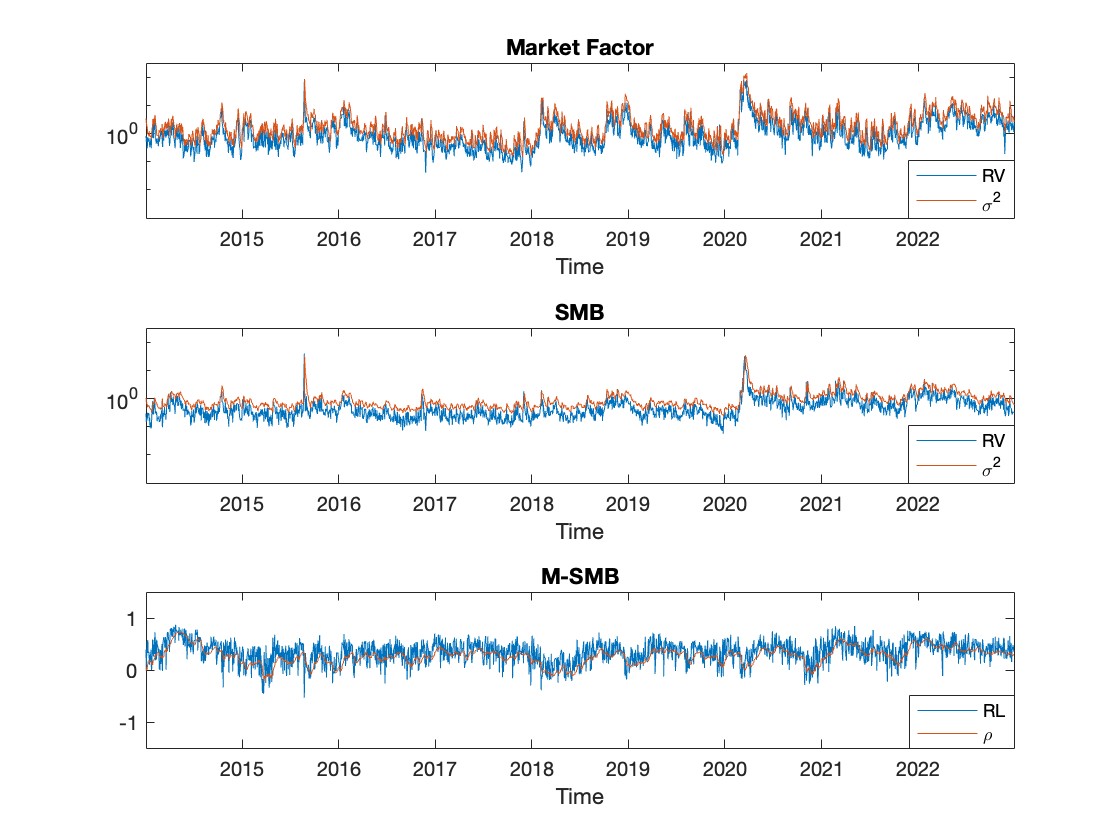}
\caption{Annualized log-transformed realized and fitted “FF-HD DCC-HEAVY'' conditional variances of the market and SMB factors and the corresponding correlations}\label{fig:4}
\end{figure}

\begin{figure}[H]
\centering
\includegraphics[height=9cm,width=\textwidth]{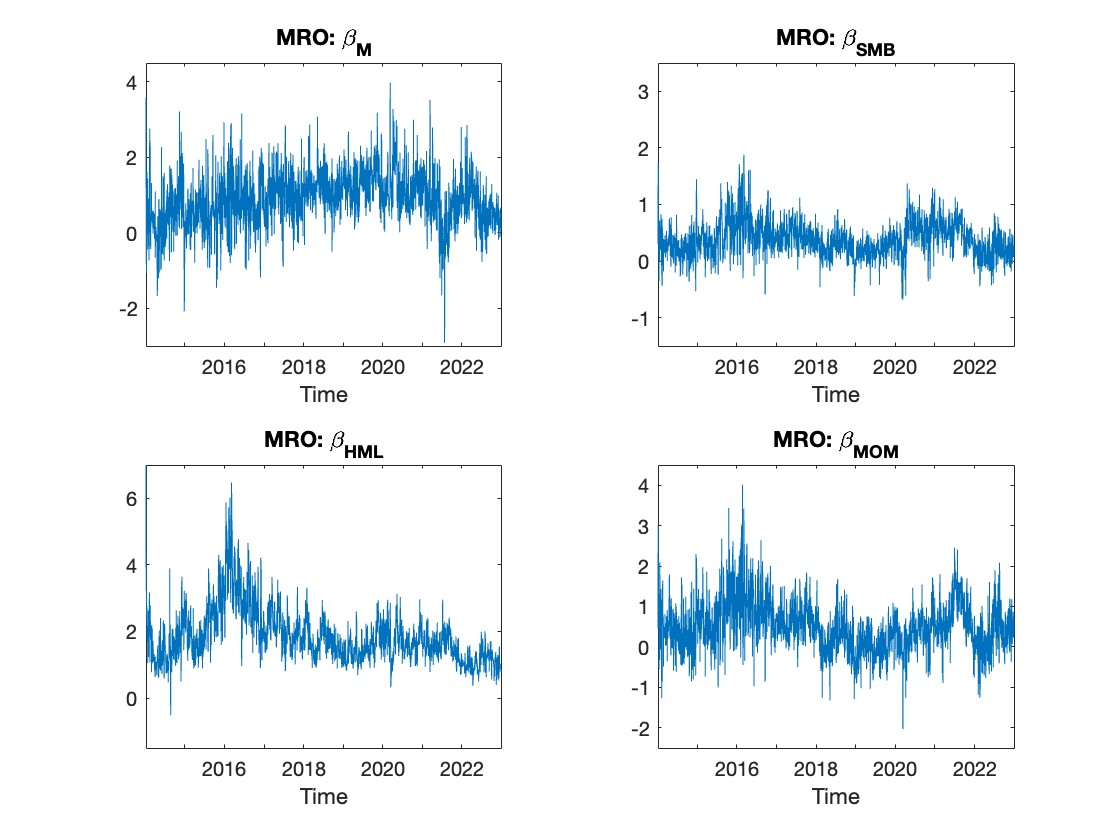}
\caption{Fitted “4F-HD DCC-HEAVY'' betas for MRO for the period 2014–2022}
\label{fig:5}
\end{figure}

\begin{table} [H] 
\begin{center}
\scalebox{0.9}{
\begin{threeparttable}
\caption{Parameter estimates of the core model for HD DCC-HEAVY variants} \label{Table:core model}
\begin{tabular}{l c c c c c c}
\hline
Coeff. & $w_h$ & $a_h^+$ & $a_h^-$ & $b_h$ & $\alpha_R$ & $\beta_R$ \\ 
Model			& & & & & & \\
\hline
\textbf{4F-HD DCC-HEAVY} & 0.000 & 0.695 & 0.772 & 0.592 & 0.051 & 0.871 \\
\textbf{FF-HD DCC-HEAVY} & 0.000 & 0.859 & 0.789 & 0.586 & 0.123 & 0.898 \\
\textbf{M-HD DCC-HEAVY} & 0.000 & 0.808 & 1.327 & 0.402 & - & - \\
\hline
\hline
Coeff.	& $w_m$ & $a_m^+$ & $a_m^-$ & $b_m$ & $\alpha_P$ & $\beta_P$ \\ 
Model			& & & & & & \\
\hline
\textbf{4F-HD DCC-HEAVY} & 0.000 & 0.181 & 0.441 & 0.377 & 0.000 & 0.998 \\
\textbf{FF-HD DCC-HEAVY} & 0.000 & 0.181 & 0.461 & 0.358 & 0.000 & 0.998 \\
\textbf{M-HD DCC-HEAVY} & 0.000 & 0.000 & 0.610 & 0.389 & - & - \\
\hline 
\hline
\end{tabular}
\begin{tablenotes}[flushleft]
\footnotesize 
\item 	
Presented are the estimates of the parameters that appear in
the HD DCC-HEAVY equations of the core model for the conditional variances and correlations (upper panel) and the corresponding realized analogues (lower panel). Columns II-V provide the average of the estimates of the univariate models for the variance of each factor. Columns VI-VII provide estimates of the parameters of the correlations. The estimation period is January 2, 2014, to December 30, 2022, i.e., \(T=2266\).
\end{tablenotes}
\end{threeparttable}}
\end{center}
\end{table}
\vspace{-0.5cm}In Figure \ref{fig:4}, we plot realized variances and correlations for the market and SMB factors against the corresponding fitted conditional measures via the “FF-HD DCC-HEAVY'' model. Clearly, each conditional variance tracks the realized series closely. The apparent downward bias of the realized measure is due to the fact that it is computed over a fraction of the day, i.e., the roughly 6.5 hours where assets are actively traded. In addition, Figure \ref{fig:4} demonstrates a significant temporal variation in the correlation dynamics of selected factors, suggesting the potential importance of defining a time-varying specification for the factor covariances.
\newpage
For HD DCC-HEAVY models, we implicitly assume that the correlations across the selected cross-section of asset returns are explained via either a single, three, or four sources of systematic risk, i.e., market, size, value, and momentum. To present the rich dynamics of estimated betas, we graphically illustrate the “4F-HD DCC-HEAVY'' fitted measures for the Marathon Oil Corporation (MRO) in Figure \ref{fig:5}. The average market beta is around 1, implying the MRO tracks the S\&P500 dynamics. Conversely, the means of the size and momentum betas lie in the interval 0.4-0.5, while the average HML beta is around 2.2. The exposure to the value risk factor varies the most. All the betas hit extreme negative values during the coronavirus pandemic. The corresponding summary statistics are given in Table \ref{table:4}.

\begin{table} [H] 
\begin{center}
\scalebox{0.9}{
\begin{threeparttable}
\caption{“4F-HD DCC-HEAVY'' beta estimates for MRO} \label{table:4}
\begin{center}
\begin{tabular}{l | c c c c | c c c c}
\hline
& $\hat{\beta}_M$ & $\hat{\beta}_{SMB}$ & $\hat{\beta}_{HML}$ & $\hat{\beta}_{MOM}$ & $\hat{\beta}_M$ & $\hat{\beta}_{SMB}$ & $\hat{\beta}_{HML}$ & $\hat{\beta}_{MOM}$ \\ 
\hline
mean & \phantom{-}1.262 & \phantom{-}0.416 &\phantom{-}2.194 &\phantom{-}0.486 &\phantom{-}1.224 & \phantom{-}0.439 &\phantom{-}1.719 &\phantom{-}0.451 \\
sd. & \phantom{-}0.754 &\phantom{-}0.302 &\phantom{-}0.921 &\phantom{-}0.603 &\phantom{-}0.774 &\phantom{-}0.320 &\phantom{-}0.502 &\phantom{-}0.579\\
&&&&&&&&\\
min & -1.613 & -0.620 & \phantom{-}0.372 & -2.059 & -1.613 & -0.620 & \phantom{-}0.536 & -2.059 \\
1 & -0.420 & -0.184 & \phantom{-}0.881 & -0.723 & -0.385 & -0.196 & \phantom{-}0.754 & -0.709\\
5 & \phantom{-}0.051 & -0.012 &\phantom{-}1.114 & -0.413 &\phantom{-}0.009 & -0.067 & \phantom{-}0.990 &\phantom{-}0.376\\
10 & \phantom{-}0.349 & \phantom{-}0.067 &\phantom{-}1.289 & -0.220  &\phantom{-}0.347 & \phantom{-}0.035 &\phantom{-}1.081 & \phantom{-}0.186\\
25 & \phantom{-}0.782 & \phantom{-}0.205 &\phantom{-}1.593 &\phantom{-}0.085 &\phantom{-}0.724 & \phantom{-}0.188 &\phantom{-}1.356 & \phantom{-}0.090 \\
50 & \phantom{-}1.239 &\phantom{-}0.390 &\phantom{-}1.981 &\phantom{-}0.435 &\phantom{-}1.183 & \phantom{-}0.445 &\phantom{-}1.713 & \phantom{-}0.379 \\
75 & \phantom{-}1.716 &\phantom{-}0.588 &\phantom{-}2.573 &\phantom{-}0.820 &\phantom{-}1.642 &\phantom{-}0.670 &\phantom{-}2.012 &\phantom{-}0.729 \\
90 & \phantom{-}2.198 &\phantom{-}0.810 &\phantom{-}3.345 &\phantom{-}1.310 &\phantom{-}2.173 &\phantom{-}0.862 &\phantom{-}2.378 &\phantom{-}1.299 \\
95 & \phantom{-}2.477 &\phantom{-}0.947 &\phantom{-}3.952 &\phantom{-}1.592 &\phantom{-}2.586 &\phantom{-}0.947 &\phantom{-}2.597 &\phantom{-}1.577 \\
99 & \phantom{-}3.162 &\phantom{-}1.239 &\phantom{-}5.333 &\phantom{-}2.198 &\phantom{-}3.382 &\phantom{-}1.169 &\phantom{-}3.085 &\phantom{-}1.875 \\
max & \phantom{-}7.536 &\phantom{-}2.406 &\phantom{-}11.182\ph &\phantom{-}3.154 &\phantom{-}4.597 &\phantom{-}1.289 &\phantom{-}3.683 &\phantom{-}2.547 \\
\hline
\hline
\end{tabular}
\begin{tablenotes}[flushleft]
\footnotesize 
\item 
Presented are the “4F-HD DCC-HEAVY'' estimates of betas for MRO for the full sample period, i.e., 2014-2022 (left panel), and the COVID turbulence, i.e., 2019-2022 (right panel). 
\end{tablenotes}
\end{center}
\end{threeparttable}}
\end{center}
\end{table}

\subsection{Out-of-Sample Forecasting}
We compute the out-of-sample forecasts discussed in Section 4\footnote{The multi-step ahead forecasts cannot be generated via the hierarchical-type Realized Beta GARCH model of \cite{hansen2014realized}.} with regard to all the asymmetric hierarchical-type factor models, which fit the data better compared to the cDCC model, i.e., “4F-HD DCC-HEAVY'', “FF-HD DCC-HEAVY'',“M-HD DCC-HEAVY'', and Realized Beta GARCH. Starting from the fitting period from January 2, 2014, to December 31, 2018 \((T_e = 1274)\), we generate the forecasts
by re-estimating the models with \(T_e\) daily observations and then producing a sequence of 1-step-ahead predictions based on the updated parameter estimates. We consider the two out-of-sample forecasting periods. The first, characterized by the relatively low volatility of returns, includes the period January 2019 - December 2019. The second lasts until the end of 2022, with the volatility at a relatively high level triggered by the COVID pandemic. In addition, we evaluate the out-of-sample monthly conditional covariance forecasts built upon `HF' daily returns of all the models detailed in Appendix \ref{AppC}.

\subsubsection{Statistical Accuracy} 
In order to assess the statistical accuracy of all models, we adopt the two loss functions that produce the consistent ranking (\cite{laurent2013loss}, \cite{patton2011volatility}), i.e., the Euclidean distance (ED) and squared Frobenius norm (FN).

The first is based on the \(\vech(\cdot)\)\footnote{The operator that stacks the
lower triangular part of a symmetric \(N\times N\) matrix argument into a \(N(N+1)/2 \times 1\) vector.} transformation of the forecast error matrix, where the prediction errors on variances and covariances are equally weighted, i.e.,
\begin{equation} \label{28}
ED_{t}(C_{t+1}, \hat{H}_{t+1}) = \vech({C}_{t+1} - \hat{H}_{t+1})'\Iden_{N*}\vech({C}_{t+1} - \hat{H}_{t+1}),
\end{equation}
where \(\hat{H}_{t+1}\) is the conditional forecast of the covariances of \(r_{t+1}\), \(C_{t+1}\) is a proxy for the unobserved covariance matrix at time \(t + 1\), and \(\Iden_{N*}\) is the identity matrix of order \(N(N+1)/2\). Indeed, the natural proxy for latent covariances is given by the RC.\footnote{Conversely, the adoption of \(r_{t}r'_{t}\) appears more suitable when forecasting monthly covariances.}

The second loss function is the matrix equivalent of the MSE loss function, where the weights on the covariance forecast errors are doubled compared to the ones on variances, i.e.,
\begin{equation} \label{29}
FN_{t}(C_{t+1}, \hat{H}_{t+1}) = \trace[(C_{t+1} - \hat{H}_{t+1})'(C_{t+1} - \hat{H}_{t+1})] = \sum_{i,j}(c_{ij,t+1} - \hat{h}_{ij,t+1})^{2}.
\end{equation}

For assessing the significance of differences in the ED and FN losses across the four models, we rely on the model confidence set (MCS) approach of \cite{hansen2011model}. The MCS identifies the model or subset of models with the best forecasting performance, given the pre-specified confidence level. It is computed at the 10\% significance level using a block bootstrap (\cite{hansen2003choosing}) with 10,000 replications and the varying block length to verify the robustness of the results.

\begin{table}[H]
\begin{center}
\scalebox{0.9}{
\begin{threeparttable}
\caption{Model confidence sets at the 90\% level of hierarchical factor models with ED and FN loss functions}\label{table:MCS_stat}
\begin{center}
\begin{tabular}{l | c c | c c| c c}
\hline
Model & ED & MCS 2019 & ED & MCS 2020-2022 & ED & MCS 2019-2022\\
\hline
\textbf{4F-HD DCC-HEAVY} & 0.023 & 0.011 & 0.560 & \textbf{0.114} & 0.434 & \textbf{0.103} \\ 
\textbf{FF-HD DCC-HEAVY} & 0.025 & 0.002 & 0.943 & 0.034 & 0.727 & 0.011 \\ 
\textbf{M-HD DCC-HEAVY} & \textbf{0.020} & \textbf{1.000} & \textbf{0.375} & \textbf{1.000} & \textbf{0.291} & \textbf{1.000}\\
\textbf{Realized Beta GARCH} & 0.090 & 0.002 & 0.764 & 0.034 & 0.624 & 0.094 \\
\hline
Model & FN & MCS 2019 & FN & MCS 2020-2022 & FN & MCS 2019-2022\\
\hline
\textbf{4F-HD DCC-HEAVY} & 0.028 & \textbf{0.315} & 0.816 & \textbf{0.106} & 0.630 & 0.098 \\
\textbf{FF-HD DCC-HEAVY} & 0.032 & 0.035 & 1.261 & 0.034 & 0.971 & 0.012 \\ 
\textbf{M-HD DCC-HEAVY} & \textbf{0.027} & \textbf{1.000} & \textbf{0.493} & \textbf{1.000} & \textbf{0.383} & \textbf{1.000}\\
\textbf{Realized Beta GARCH} & 0.155 & 0.017 & 1.142 & 0.086 & 0.943& 0.080\\
\hline
\hline
\end{tabular}
\begin{tablenotes}[flushleft]
\footnotesize 
\item 
`ED/FN' columns: the average annualized value of ED/FN losses over the corresponding forecast period; bold values identify the minimum loss over the four models. \\
`MCS 2019' column: $p$-values of the MCS tests over the out-of-sample period, including the year 2019; bold values identify the models included in the MCS at the 90\% confidence level (i.e., $p$-values larger than 0.10). \\
`MCS 2020-2022' column: the analogous results for the period 2020-2022.\\
`MCS 2020-2022' column: the analogous results for the full out-of-sample period.\vspace{-0.5cm}
\end{tablenotes}
\end{center}
\end{threeparttable}}
\end{center}
\end{table}

Table \ref{table:MCS_stat} reports the model confidence sets at the 90\% confidence level using the ED and FN loss functions. The “M-HD DCC-HEAVY'' and “4F-HD DCC-HEAVY'' models statistically outperform the other models during both forecasting subsamples.

Considering the full out-of-sample period, only the “M-HD DCC-HEAVY'' model enters the MCS in terms of the FN loss, while the “4F-HD DCC-HEAVY'' model is additionally selected in terms of the ED. The hierarchical model of \cite{hansen2014realized} is always excluded from the reported model confidence sets with respect to both daily and monthly covariance forecasts (see Table \ref{table:MCS_stat} \& Appendix \ref{AppC}, Table \ref{table:C1}).

\subsubsection{Economic Performance}
In order to perform an economic evaluation of the forecasting performance, we rely on the global minimum variance portfolio (GMVP) optimization (e.g., \cite{BauwensXu2022HEAVY}, \cite{engle2012dynamic}) since it does not require the estimation of expected returns, providing an essentially clean framework for assessing the merits of distinct covariance forecasting models.

In this regard, given a covariance matrix forecast \(\hat{H}_{t+1}\), the portfolio weights \(\hat{\omega}_{t+1}\) are obtained by solving the minimization problem:
\begin{equation} \label{30}
\underset{\omega_{t+1}}{\min}\hspace{0.1cm} \omega_{t+1}'\hat{H}_{t+1}\omega_{t+1} \quad \st \quad \omega_{t+1}'\textbf{1}=1,
\end{equation}
where \textbf{1} is a \(\textit{N} \times 1\) vector of ones.

It follows readily that the optimal GMVP weights are:
\begin{equation} \label{31}
\hat{\omega}_{t+1} = \frac{\hat{H}_{t+1}^{-1}\textbf{1}}{\textbf{1}'\hat{H}_{t+1}^{-1}\textbf{1}}.
\end{equation}

Given the main aim of assessing the accuracy of distinct covariance matrix estimators, our performance measures do not take into account transaction costs. In addition, we consider the optimization under a short-selling restriction and compute the weights via numerical optimization, i.e., MATLAB Financial Toolbox, given the absence of a closed-form analytical solution.
\newpage
Accordingly, we adopt the MCS to select the best-performing models that minimize the standard deviation (SD) of the portfolios obtained by applying the computed weights to the observed returns.

The results presented in Table \ref{table:MCS_econ} show that the “4F-HD DCC-HEAVY'' model not only provides for the lowest out-of-sample SD during the calm period but also enters the MCS when the volatility is at a relatively high level. Correspondingly, considering the entire out-of-sample period, the MCS includes only the “4F-HD DCC-HEAVY'' model, while it also outperforms the other models with respect to both long-only portfolios (Appendix \ref{AppB}, Table \ref{table:B1}) and monthly forecasts (Appendix \ref{AppC}, Table \ref{table:C2}). Therefore, in contrast to the statistical performance where the “M-HD DCC-HEAVY'' model is jointly superior with the “4F-HD DCC-HEAVY'', only the latter appears preferable from a variance minimization perspective.

\begin{table}[H]
\begin{center}
\scalebox{0.9}{
\begin{threeparttable}
\caption{Model confidence sets at the 90\% level of hierarchical factor models, with GMVP loss function} \label{table:MCS_econ}
\begin{center}
\begin{tabular}{l | c c | c c | c c}
\hline
Model & SD & MCS 2019 & SD & MCS 2020-2022 & SD & MCS 2019-2022 \\
\hline
\textbf{4F-HD DCC-HEAVY} & \textbf{0.556} & \textbf{1.000} & \textbf{0.779} & \textbf{1.000} & \textbf{0.727} & \textbf{1.000} \\ 
\textbf{FF-HD DCC-HEAVY} & 0.585 & 0.011 & 0.789 & \textbf{0.313} & 0.741 & 0.086\\ 
\textbf{M-HD DCC-HEAVY} & 0.653 & 0.000 & 0.870 & 0.000 & 0.818 & 0.000 \\
\textbf{Realized Beta GARCH} & 0.650 & 0.000 & 0.868 & 0.000  & 0.816 & 0.000\\
\hline
\hline
\end{tabular}
\begin{tablenotes}[flushleft]
\footnotesize 
\item 
`SD' columns: the average annualized standard deviation of GMVP returns over the corresponding forecast period; bold values identify the minimum loss over the four models. \\
`MCS 2019' column: $p$-values of the MCS tests over the out-of-sample period, including the year 2019; bold values identify the models included in the MCS at the 90\% confidence level  (i.e., $p$-values larger than 0.10). \\
`MCS 2020-2022' column: the analogous results for the period 2020-2022. \\
'MCS 2019-2022' column: the analogous results for the full out-of-sample period.
\end{tablenotes}
\end{center}
\end{threeparttable}}
\end{center}
\end{table}

\vspace{-0.5cm}We additionally examine some basic features of the portfolios, including the average return (AR), i.e., the average of out-of-sample returns for the corresponding period; the information ratio (IR), i.e., the ratio AR/SD; portfolio turnover rates (TO); and the proportion of short positions (SP).

The latter two are specified as follows: 
\begin{equation} \label{32}
TO_t = \sum_{i}^{N}\left\lvert\hat{w}_{t}^{i} - \hat{w}_{t-1}^{i}\frac{1+r_{t-1}^{i}}{1+r_{t-1}^{p}}\right\rvert;
\end{equation}
\begin{equation} 
\label{33}
SP_t = \sum_{i}^{N}\1_{\{\hat{w}_{t}^{i}<0\}},
\end{equation}
where \(r_{t}^{p}\) is the total return of the portfolio for the day \(t\), \(\hat{w}_{t}^{i}\) and \(r_{t}^{i}\) are the weight and return of stock \(i\), respectively, and \(\1_{\{\cdot\}}\) denotes the indicator function.\footnote{We do not set constraints on the turnover and leverage proportion in the optimization.}

The results reported in Table \ref{table:B2} further confirm that hierarchical HD DCC-HEAVY models consistently outperform the Realized Beta GARCH model (\cite{hansen2014realized}). In particular, the “FF-HD DCC-HEAVY'' features the highest IR during turbulent periods and overall. On the other hand, the propensity of models with respect to short positions is very similar, and the increasing trend of the average monthly turnover rates for all models is also visible during turmoils.

Finally, to assess the economic gains of utilizing distinct HD DCC-HEAVY covariance matrix estimators, following \cite{fleming2001economic,fleming2003economic}, we determine the maximum performance fee a risk-averse investor would be willing to pay to switch from using one model to another.\footnote{We report the results computed with non-negatively weighted portfolios since short-selling is difficult to implement in practice.} Accordingly, we assume that the investor has quadratic preferences of the form:
\begin{equation} \label{34}
U(r_t^p) = 1+r_t^p-\frac{\gamma}{2(1+\gamma)}(1+r_t^p)^2,
\end{equation}
where \(r_t^p\) is the portfolio return and \(\gamma\) is the investor’s relative risk aversion, taking values 1 and 10 (\cite{fleming2003economic}). We determine a fee \(\Delta_\gamma\) by equating the average realized utilities from two alternative portfolios, i.e.,
\begin{equation} \label{35}
\sum_{t=1}^T{U(r_t^{p_1})} = \sum_{t=1}^T{U(r_t^{p_2}-\Delta_\gamma)},
\end{equation}
where \(r_t^{p_1}\) and \(r_t^{p_2}\) are the portfolio returns related to competing HD DCC-HEAVY forecasting strategies.
\newpage
Major observations based on the results in Table \ref{table:7} are as follows: First, by utilizing the “4F-HD DCC-HEAVY'' covariance forecasts, a risk-averse investor can achieve economic gains that become more pronounced during the crisis period. In particular, an investor with low (high) risk aversion would be willing to pay on average 6 (8) bps to switch from the “FF-HD DCC-HEAVY'' to the “4F-HD DCC-HEAVY'' strategy and around 1 bps for switching from the “M-HD DCC-HEAVY''. These results provide further support that the “4F-HD DCC-HEAVY'' might be a preferable hierarchical model from the investor's point of view.
\vspace{0.5cm}
\begin{table}[H]
\begin{center}
\scalebox{0.9}{
\begin{threeparttable}
\caption{BPS fees for switching from simpler HD DCC-HEAVY to the “4F-HD DCC-HEAVY'' covariance matrix forecasts} \label{table:7}
\begin{tabular}{l | c c | c c | c c}
\hline
Period & \multicolumn{2}{c}{2020-2022} & \multicolumn{2}{c}{2019-2022} \\ 
\hline
Model & $\Delta_{1}$ & $\Delta_{10}$ & $\Delta_{1}$ & $\Delta_{10}$ \\
\hline
\textbf{FF-HD DCC-HEAVY} & 6.29 & 8.03 & 5.20 & 6.52 \\ 
\textbf{M-HD DCC-HEAVY} & 0.97 & 0.97 & 0.31 & 0.32 \\    
\hline
\hline
\end{tabular}
\begin{tablenotes}[flushleft]
\footnotesize 
\item 
`$\Delta_{\gamma}$' columns: the basis points fee an investor with quadratic utility and relative risk aversion $\gamma$ would pay to switch from the covariance matrix estimator indicated in column 1 to the “4F-HD DCC-HEAVY'' over the period indicated in row 1.
\end{tablenotes}
\end{threeparttable}}
\end{center}
\end{table}

\vspace{-1.5cm}\section{Conclusion}
In this paper, we introduce a class of models for high-dimensional covariance matrices by combining the hierarchical factor modelling approach of \cite{hansen2014realized} that is independent from the cross-sectional series dimension and the dynamic conditional correlation formulation of a HEAVY model (\cite{noureldin2012multivariate}) recently proposed by \cite{BauwensXu2022HEAVY}. The illustrative empirical studies at distinct frequencies for the S\&P500 constituents over the period from January 1962 until January 2023 show that our method significantly outperforms the benchmark cDCC model and existing hierarchical factor model in statistical and economic terms. The results are robust under different market conditions. 

As such, we confirm the findings of, e.g., \cite{noureldin2012multivariate}, \cite{gorgi2019realized}, \cite{BauwensXu2022HEAVY}, etc., that utilizing the HEAVY-type models based on more accurate realized measurements of covariances than low-frequency data-based GARCH-type models improves the forecasts of the conditional covariance matrix of daily returns. In addition, the adopted DCC structure (see e.g., \cite{corsi2021dcc}) and/or the rich asymmetric dynamics could be important since our models readily outperform the Realized Beta GARCH model of \cite{hansen2014realized}.

The first avenue for future research concerns the development of the 
testing procedures to verify the relevance of the factor set and thus adopt the optimal HD DCC-HEAVY model. Second, many
other observable factors can be explored, e.g., RMW (Robust Minus Weak), CMA (Conservative Minus Aggressive) (\cite{fama2015five}), etc., but also the intrinsic time-varying betas.
\vspace{-0.5cm}

\bibliographystyle{apalike}
\begin{singlespace}
\bibliography{main.bbl}
\end{singlespace}
\newpage

\appendix
\section*{Appendices}
\renewcommand{\thesubsection}{\Alph{subsection}}
\renewcommand{\thetable}{A\arabic{table}}
\setcounter{table}{0}
\subsection{Summary Statistics for Individual Assets}\label{AppA} 
\vspace{0.25cm}

The individual stocks for the empirical analyses include: American Electric Power Company, Inc. (AEP), The Boeing Company (BA), Caterpillar Inc. (CAT), Chevron Corporation (CVX), DTE Energy Company (DTE), Consolidated Edison, Inc. (ED), General Dynamics Corporation (GD), Honeywell International Inc. (HON), International Business Machines Corporation (IBM), International Paper Company (IP), The Coca-Cola Company (KO), The Kroger Co. (KR), 3M Company (MMM), Altria Group, Inc. (MO), Merck \& Co., Inc. (MRK), Marathon Oil Corporation (MRO), Motorola Solutions, Inc. (MSI), The Procter \& Gamble Company (PG), and Exxon Mobil Corporation (XOM).
\vspace{0.25cm}
\begin{table} [H] 
\begin{adjustwidth}{-.8in}{-.8in} 
\begin{center}
\scalebox{0.8}{
\begin{threeparttable}
\caption{Summary statistics of daily measures for individual assets} \label{table:A1}
\begin{tabular}{l|ll|ll|ll|c}
 \hline
 &\ph\ph\ph\ph \textit{r}$_{cc}^2$ &\ph\ph\ph\ph $RV$ &\ph\ph\ph\ph $P$ &\ph\ph\ph\ph $N$ &\ph\ph\ph $GJR_{P}$ &\ph\ph\ph $GJR_{N}$ & $RL$ \\
 \hline
AEP & \ph4.31 (15.87) & \ph3.83 (11.08) & \ph1.92 \ph(6.16) & \ph1.91 \ph(5.23) & \ph1.81 \ph(4.68) & \ph2.02 (10.40) & 0.21 (0.24) \\
BA & 16.37 (75.43) & \ph9.45 (29.80) & \ph4.65 (14.06) & \ph4.80 (16.17) & \ph4.21 (14.33) & \ph5.24 (26.96) & 0.50 (0.21) \\
CAT & \ph8.67 (22.76) & \ph5.82 \ph(8.82) & \ph2.93 \ph(5.12) & \ph2.89 \ph(4.09) & \ph2.84 \ph(5.88) & \ph2.98 \ph(7.76) & 0.53 (0.19) \\
CVX & \ph9.19 (49.44) & \ph5.70 (12.99) & \ph2.90 \ph(7.10) & \ph2.80 \ph(6.30) & \ph2.96 (10.64) & \ph2.74 \ph(8.48) & 0.44 (0.20) \\
DTE & \ph5.14 (24.66) & \ph3.91 (11.03) & \ph1.95 \ph(5.80) & \ph1.96 \ph(5.68) & \ph1.90 \ph(6.07) & \ph2.01 \ph(9.62) & 0.22 (0.24) \\
ED & \ph4.43 (21.82) & \ph3.70 \ph(8.62) & \ph1.86 \ph(4.96) & \ph1.84 \ph(4.07) & \ph1.81 \ph(5.22) & \ph1.89 \ph(7.35) & 0.17 (0.24) \\
GD & \ph5.17 (14.95) & \ph4.11 \ph(7.71) & \ph2.10 \ph(4.34) & \ph2.01 \ph(3.65) & \ph1.97 \ph(4.40) & \ph2.14 \ph(6.97) & 0.46 (0.22) \\
HON & \ph5.31 (20.16) & \ph3.77 (10.08) & \ph1.88 \ph(5.21) & \ph1.89 \ph(5.10) & \ph1.70 \ph(4.45) & \ph2.07 \ph(9.42) & 0.57 (0.21) \\
IBM & \ph5.79 (20.44) & \ph3.40 \ph(7.11) & \ph1.73 \ph(3.91) & \ph1.67 \ph(3.40) & \ph1.60 \ph(4.24) & \ph1.80 \ph(6.20) & 0.50 (0.22) \\
IP & \ph8.34 (26.73) & \ph6.17 (12.77) & \ph3.11 \ph(7.01) & \ph3.06 \ph(6.32) & \ph2.85 \ph(6.94) & \ph3.32 (11.57) & 0.44 (0.22) \\
KO & \ph3.41 (12.19) & \ph2.67 \ph(7.24) & \ph1.33 \ph(3.70) & \ph1.34 \ph(3.77) & \ph1.24 \ph(3.70) & \ph1.43 \ph(6.50) & 0.34 (0.24) \\
KR & \ph8.51 (33.01) & \ph6.57 (13.69) & \ph3.37 \ph(8.47) & \ph3.20 \ph(6.73) & \ph3.33 (10.69) & \ph3.24 \ph(9.72) & 0.28 (0.22) \\
MMM & \ph5.19 (18.88) & \ph3.47 \ph(7.01) & \ph1.69 \ph(3.45) & \ph1.78 \ph(3.87) & \ph1.66 \ph(5.05) & \ph1.81 \ph(5.44) & 0.51 (0.23) \\
MO & \ph5.20 (16.17) & \ph4.37 (11.93) & \ph2.15 \ph(6.09) & \ph2.22 \ph(6.46) & \ph1.93 \ph(4.73) & \ph2.44 (11.38) & 0.30 (0.23) \\
MRK & \ph4.69 (14.29) & \ph3.63 \ph(6.47) & \ph1.83 \ph(3.57) & \ph1.80 \ph(3.11) & \ph1.75 \ph(3.74) & \ph1.88 \ph(5.89) & 0.38 (0.23) \\
MRO & 31.28 (220.6) & 21.64 (44.65) & 10.76 (24.54) & 10.88 (22.20) & 10.79 (30.78) & 10.85 (35.78) & 0.39 (0.19) \\
MSI & \ph6.41 (21.79) & \ph4.47 (10.06) & \ph2.24 \ph(4.92) & \ph2.23 \ph(5.53) & \ph2.12 \ph(5.33) & \ph2.35 \ph(9.10) & 0.45 (0.21) \\
PG & \ph3.44 (13.10) & \ph2.91 \ph(8.87) & \ph1.51 \ph(5.24) & \ph1.40 \ph(3.83) & \ph1.32 \ph(4.12) & \ph1.59 \ph(8.11) & 0.31 (0.25) \\
XOM & \ph7.77 (23.52) & \ph5.30 (11.12) & \ph2.69 \ph(5.97) & \ph2.61 \ph(5.48) & \ph2.61 \ph(8.36) & \ph2.69 \ph(8.24) & 0.44 (0.21) \\
\hline
\end{tabular}
\begin{tablenotes}[flushleft]
\footnotesize 
\item 
$r_{cc}^2$: squared close-to-close return;
$RV$: realized variance; $P$: positive semi-variance; $N$: negative semi-variance;
$GJR_P$: $RV$ if daily return is positive, $0$ if negative; 
$GJR_N$:  $RV$ if daily return is negative, $0$ if positive;
$RL$:  realized correlation, the average and sd of realized correlations with the market factor.
\end{tablenotes}
\end{threeparttable}}
\end{center}
\end{adjustwidth}
\end{table}

\newpage
\renewcommand{\thetable}{B\arabic{table}}
\setcounter{table}{0}
\subsection{Out-of-Sample Performance cont’d}\label{AppB} 
\vspace{0.25cm}
\vspace{-0.5cm}\begin{table}[H]
\begin{center}
\scalebox{0.8}{
\begin{threeparttable}
\caption{Model confidence set at the 90\% level of hierarchical factor models with GMVP loss function under long-only portfolios} \label{table:B1}
\begin{center}
\begin{tabular}{l | c c | c c | c c}
\hline
Period & \multicolumn{2}{c}{2019} & \multicolumn{2}{c}{2020-2022} & \multicolumn{2}{c}{2019-2022} \\
\hline
Model & SD & MCS & SD & MCS & SD & MCS \\
\hline
\textbf{4F-HD DCC-HEAVY} & 0.130 & 0.003 & \textbf{0.198} & \textbf{1.000}  & \textbf{0.182} & \textbf{1.000} \\
\textbf{FF-HD DCC-HEAVY} & 0.131 & 0.002 & 0.208 & 0.012 & 0.191 & 0.007 \\ 
\textbf{M-HD DCC-HEAVY} & 0.130 & 0.002 & 0.203 & 0.037 & 0.186 & 0.011 \\
\textbf{Realized Beta GARCH} & \textbf{0.125} & \textbf{1.000} & 0.201 & \textbf{0.513} & 0.183 & \textbf{0.787} \\
\hline
\hline
\end{tabular}
\begin{tablenotes}[flushleft]
\footnotesize 
\item 
`SD' columns: the average annualized standard deviation of GMVP returns with short sale restrictions over the forecast period indicated in row 1; bold values identify the minimum loss over the four models. \\
`MCS' columns: $p$-values of the MCS tests over the out-of-sample period indicated in row 1; bold values identify the models included in the MCS at the 90\% confidence level  (i.e., $p$-values larger than 0.10). \\
\end{tablenotes}
\end{center}
\end{threeparttable}}
\end{center}
\end{table}

\vspace{-0.5cm}\begin{table}[H]
\begin{center}
\scalebox{0.8}{
\begin{threeparttable}
\caption{Alternative economic performance measures for hierarchical factor models based on GMVP optimization} \label{table:B2}
\begin{tabular}{l | c c | c c | c c | c c | c c}
\hline
Period & \multicolumn{2}{c}{2020-2022} & \multicolumn{2}{c}{2019-2022} & \multicolumn{2}{c}{2020-2022} & \multicolumn{2}{c}{2019-2022}\\ 
\hline
Model & AR & IR & AR & IR & TO & SP & TO & SP\\
\hline
\textbf{4F-HD DCC-HEAVY} & 0.152 & 0.196 & 0.102 & 0.140&0.964&0.476& 0.951 & 0.477\\ 
\textbf{FF-HD DCC-HEAVY} & 0.155 & 0.196 & 0.132 & 0.178&1.008&0.476& 0.998&0.476\\   
\textbf{M-HD DCC-HEAVY} & 0.108 & 0.124 & 0.083 & 0.101&1.027&0.474&1.019&0.474 \\
\textbf{Realized Beta GARCH} & 0.083 & 0.096 & 0.096 & 0.118&0.992&0.469&0.985&0.467\\
\hline
\hline
\end{tabular}
\begin{tablenotes}[flushleft]
\footnotesize 
\item 
`AR' columns: the average annualized GMVP return over the forecast period indicated in row 1. \\
`IR' columns: the average annualized AR/SD ratio over the forecast period indicated in row 1. \\
`TO' columns: the average portfolio turnover over the forecast period indicated in row 1. \\
`SP' columns: the average leverage proportion over the forecast period indicated in row 1. \\
\end{tablenotes}
\end{threeparttable}}
\end{center}
\end{table}

\renewcommand{\thetable}{C\arabic{table}}
\setcounter{table}{0}
\subsection{Out-of-Sample Performance at Daily vs. Monthly Frequency}\label{AppC} 
\vspace{0.25cm}
We start from the fitting period from January 1962 to December 2016 \((T_e = 660)\) and re-estimate the models every year on a rolling window with \(T_e\) observations in order to generate a sequence of 1-step-ahead monthly covariance predictions. Again, the two out-of-sample periods are analyzed. I.e., the first is characterized by the relatively low volatility of returns, including the years 2017-2019. The second coincides with the intra-daily vs. daily analyses, lasting until the end of 2022. We also report the results for a full out-of-sample period.
\vspace{1cm}
\begin{table}[H]
\begin{center}
\scalebox{0.8}{
\begin{threeparttable}
\caption{Model confidence sets at the 90\% level of hierarchical factor models with ED and FN loss functions} \label{table:C1}
\begin{center}
\begin{tabular}{l | c c | c c| c c}
\hline
Model & ED & MCS 2017-2019 & ED & MCS 2020-2022 & ED & MCS 2017-2022\\
\hline
\textbf{4F-HD DCC-HEAVY} & 0.080 & 0.006 & 0.828 & 0.002& 0.454 & 0.005 \\ 
\textbf{FF-HD DCC-HEAVY} & \textbf{0.075} & \textbf{1.000} & \textbf{0.796} & \textbf{1.000} & \textbf{0.435} & \textbf{1.000} \\ 
\textbf{M-HD DCC-HEAVY} & 0.075 & \textbf{0.885} & 0.879 & 0.002 & 0.477 & 0.005 \\
\textbf{Realized Beta GARCH} & 0.078 & 0.091 & 0.951 & 0.002 & 0.515 & 0.005 \\
\hline
Model & FN & MCS 2017-2019 & FN & MCS 2020-2022 & FN & MCS 2017-2022\\
\hline
\textbf{4F-HD DCC-HEAVY} & 0.138 & 0.002 & 1.340 & 0.002 & 0.739 & 0.003 \\
\textbf{FF-HD DCC-HEAVY} & 0.132 & \textbf{0.320} & \textbf{1.276} & \textbf{1.000} & \textbf{0.704} & \textbf{1.000}\\ 
\textbf{M-HD DCC-HEAVY} & \textbf{0.129} & \textbf{1.000} & 1.399 & 0.002 & 0.764 & 0.003 \\
\textbf{Realized Beta GARCH} & 0.135 & 0.008 & 1.514 & 0.002 & 0.824 & 0.003 \\ 
\hline
\hline
\end{tabular}
\begin{tablenotes}[flushleft]
\footnotesize 
\item 
`ED/FN' columns: the average annualized value of ED/FN losses over the corresponding forecast period; bold values identify the minimum loss over the four models. \\
`MCS 2017-2019' column: $p$-values of the MCS tests over the out-of-sample period, including the years 2017-2019; bold values identify the models included in the MCS at the 90\% confidence level  (i.e., $p$-values larger than 0.10). \\
`MCS 2020-2022' column: the analogous results for the period 2020-2022.\\
`MCS 2017-2022' column: the analogous results for the full out-of-sample period 2017-2022.\vspace{-0.5cm}
\end{tablenotes}
\end{center}
\end{threeparttable}}
\end{center}
\end{table}

\begin{table}[H]
\begin{center}
\scalebox{0.8}{
\begin{threeparttable}
\caption{Model confidence sets at the 90\% level of hierarchical factor models with GMVP loss function} \label{table:C2}
\begin{center}
\begin{tabular}{l | c c | c c| c c}
\hline
Model & SD & MCS 2017-2019 & SD & MCS 2020-2022 & SD & MCS 2017-2022\\
\hline
\textbf{4F-HD DCC-HEAVY} & 0.653 & 0.098 & \textbf{0.571} & \textbf{1.000} & \textbf{0.612} & \textbf{1.000}\\ 
\textbf{FF-HD DCC-HEAVY} & 0.715 & 0.000 & 0.709 & 0.088 & 0.712 & 0.027 \\ 
\textbf{M-HD DCC-HEAVY} & \textbf{0.565} & \textbf{1.000} & 0.952 & 0.000 & 0.758 & 0.005\\
\textbf{Realized Beta GARCH} & 0.643 & 0.098 & 0.878 & 0.000  & 0.761 & 0.005\\
\hline
\hline
\end{tabular}
\begin{tablenotes}[flushleft]
\footnotesize 
\item 
`SD' columns: the average annualized standard deviation of GMVP returns over the corresponding forecast period; bold values identify the minimum loss over the four models. \\
`MCS 2017-2019' column: $p$-values of the MCS tests over the out-of-sample period, including the years 2017-2019; bold values identify the models included in the MCS at the 90\% confidence level  (i.e., $p$-values larger than 0.10). \\
`MCS 2020-2022' column: the analogous results for the period 2020-2022.\\
`MCS 2017-2022' column: the analogous results for the full out-of-sample period 2017-2022.\vspace{-0.5cm}
\end{tablenotes}
\end{center}
\end{threeparttable}}
\end{center}
\end{table}

\end{document}